\documentclass[floats,floatfix,amssymb,prd,twocolumn,superscriptaddress,nofootinbib]{revtex4-1}

\usepackage{subcaption}
\usepackage{ragged2e}
\DeclareCaptionJustification{justified}{\justifying}
\makeatletter
\newcommand{\subsetsim}{\mathrel{\mathpalette\subset@sim\relax}}
\newcommand{\subset@sim}[2]{%
  \vtop{\offinterlineskip\m@th
    \ialign{\hfil##\cr
      $#1\subset$\cr\noalign{\kern0.5pt}\scalebox{0.9}{$#1\sim$}\cr
    }%
  }%
}
\makeatother

\usepackage{amssymb,amsmath,verbatim,mathtools,needspace,enumitem,etoolbox,graphicx,physics,microtype,afterpage,bm}
\usepackage[dvipsnames, usenames]{xcolor}
\definecolor{linkcolor}{rgb}{0.0,0.3,0.5}
\usepackage{booktabs}

\definecolor{oucrimsonred}{rgb}{0.6, 0.0, 0.0}
\definecolor{persianblue}{rgb}{0.11, 0.22, 0.73}
\definecolor{forestgreen}{rgb}{0.13,0.35,0.13}
\usepackage[unicode, 
colorlinks=true, 
linkcolor=persianblue, 
citecolor=forestgreen, 
filecolor=persianblue,
urlcolor=oucrimsonred, 
pdfusetitle]{hyperref}

\usepackage[all]{hypcap}
\usepackage[T1]{fontenc}
\usepackage[utf8]{inputenc}
\usepackage{tabularx}
\usepackage{adjustbox}
\usepackage{float}
\usepackage{ulem}
\usepackage{xfrac}

\usepackage{multirow}
\usepackage{pifont}
\usepackage{fontawesome}
\usepackage{lmodern}

\usepackage{multirow}

\allowdisplaybreaks
\usepackage{tikz}
\usepackage{color}
\usepackage{framed}

\definecolor{rossos}{cmyk}{0,1,1,0.55}
\definecolor{bluscuro}{rgb}{0.15, 0.2, .85}
\definecolor{bluchiaro}{cmyk}{1,.3,0.,0.1}
\definecolor{ForestGreen}{rgb}{0.13, 0.55, 0.13}

\newcommand{\apjl}{Astrophys. J. Lett.}
\newcommand{\apjs}{Astrophys. J. Suppl.}

\newcommand{\aapr}{Astron. Astrophys. Rev.}

\newcommand{\jcap}{J. Cosmol. Astropart. Phys.}

\def\bea{\begin{eqnarray}}
\def\eea{\end{eqnarray}}

\def\d{{\mathrm{d}}}

\newcommand{\bs}{\begin{subequations}}
\newcommand{\es}{\end{subequations}}

\newcommand{\be}{\begin{equation}}
\newcommand{\ee}{\end{equation}}
\renewcommand{\d}{{\rm d}}

\newcommand{\llp}{\left [}
\newcommand{\rrp}{\right ]}
\newcommand{\lp}{\left (}
\newcommand{\rp}{\right )}

\def\lsim{\mathrel{\rlap{\lower4pt\hbox{\hskip0.5pt$\sim$}}
    \raise1pt\hbox{$<$}}}         
\def\gsim{\mathrel{\rlap{\lower4pt\hbox{\hskip0.5pt$\sim$}}
    \raise1pt\hbox{$>$}}}         

\newcommand{\step}{\theta_\mathrm{H}}

\newcommand{\diff}{\ensuremath{\mathrm{d}}}
\newcommand{\e}{\mathrm{e}}

\newcommand{\p}{\prime}

\newcommand{\dc}{\delta_\mathrm{c}}
\newcommand{\af}{a_\mathrm{f}}

\newcommand{\aeq}{a_\mathrm{eq}}

\newcommand{\keq}{k_\mathrm{eq}}
\newcommand{\Meq}{M_\mathrm{eq}}

\newcommand{\rhoM}{\rho_\mathrm{m}}
\newcommand{\rhoMo}{\rho_{\mathrm{m},0}}

\newcommand{\rhoRo}{\rho_{\mathrm{r},0}}

\newcommand{\rhos}{\rho_\mathrm{h}}
\newcommand{\rs}{r_\mathrm{h}}

\newcommand{\tE}{t_\text{\tiny E}}
\newcommand{\rE}{R_\text{\tiny E}}

\newcommand{\Rstar }{R_\star}

\def\chg#1{#1}

\makeatletter
\def\l@subsubsection#1#2{}
\makeatother

\newcommand{\sapienza}{Dipartimento di Fisica, Sapienza Università 
	di Roma, Piazzale Aldo Moro 5, 00185, Roma, Italy}
\newcommand{\infn}{INFN, Sezione di Roma, Piazzale Aldo Moro 2, 00185, Roma, Italy}
\newcommand{\maxplanck}{Max Planck Institute for Astrophysics, Karl-Schwarzschild-Str. 1, 85748 Garching, Germany}	
	
\begin{document}

\title{
Lensing constraints on
ultradense dark matter halos
%
}

\author{M. Sten Delos}
\email{sten@mpa-garching.mpg.de}
\affiliation{\maxplanck}

\author{Gabriele Franciolini}
\email{gabriele.franciolini@uniroma1.it}
\affiliation{\sapienza}
\affiliation{\infn}

\begin{abstract}
Cosmological observations precisely measure primordial variations in the density of the Universe at megaparsec and larger scales, but much smaller scales remain poorly constrained.
However, sufficiently large initial perturbations at small scales can lead to an abundance of ultradense dark matter minihalos that form during the radiation epoch and survive into the late-time Universe. Because of their early formation, these objects can be compact enough to produce detectable microlensing signatures. 
We investigate whether the EROS, OGLE, and HSC surveys can probe these halos by fully accounting for finite source size and extended lens effects.
We find that current data may already constrain the amplitudes of primordial curvature perturbations in a new region of parameter space, but this conclusion is strongly sensitive to yet undetermined details about the internal structures of these ultradense halos.
Under optimistic assumptions, current and future HSC data would constrain a power spectrum that features an enhancement at scales $k \sim  10^7/{\rm Mpc}$, and an amplitude as low as $\mathcal{P}_\zeta\simeq 10^{-4}$ may be accessible.
This is a particularly interesting regime because it connects to primordial black hole formation in a portion of the LIGO/Virgo/Kagra mass range 
and the production of scalar-induced gravitational waves in the nanohertz frequency range reachable by pulsar timing arrays.
\chg{These prospects motivate further study of the ultradense halo formation scenario to clarify their internal structures.}
\end{abstract}

\maketitle

\normalem

\section{Introduction}

The spectrum of primordial curvature perturbations
on large scales has been precisely constrained by a variety of observations, including the cosmic microwave background (CMB) \cite{Planck:2018vyg}, the
Lyman-alpha forest \cite{Chabanier:2019eai}, the UV galaxy luminosity function \cite{Sabti:2021unj}, and strong gravitational lensing \cite{Gilman:2021gkj}. 
These measurements bring information about the energy content of our Universe and allow us 
to constrain models of inflation (see for example Ref.~\cite{Planck:2018jri}). 
While current data are only able to constrain scales above roughly a megaparsec, 
an increasing number of new probes have been proposed to constrain primordial perturbations at smaller scales. Among these probes,
the formation of primordial black holes (PBHs) \cite{Sasaki:2018dmp,Green:2020jor,Franciolini:2021nvv}
and nonlinearly induced stochastic gravitational waves (GWs)
\cite{Domenech:2021ztg} are of particular recent interest.

PBHs can originate from the collapse 
of extreme inhomogeneities deep within the  radiation-dominated era~\cite{Zeldovich:1967lct,Hawking:1974rv,Chapline:1975ojl,Carr:1975qj} and can attain a wide range of masses~\cite{Ivanov:1994pa,GarciaBellido:1996qt,Ivanov:1997ia,Blinnikov:2016bxu}. 
Various observational bounds have been set on their abundance~\cite{Carr:2020gox}. However, 
PBHs in the asteroid-mass range could explain the dark matter, while much larger PBHs could be the seeds of supermassive black holes 
at high redshift~\cite{2010A&ARv..18..279V,Carr:2018rid,Clesse:2015wea,Serpico:2020ehh} (which may be challenging to explain 
within astrophysical formation scenarios~\cite{Volonteri:2021sfo})
and/or be responsible for a fraction of the black hole merger events 
already discovered by LIGO/Virgo/Kagra (LVK) detectors~(e.g. \cite{Bird:2016dcv,Sasaki:2016jop, Ali-Haimoud:2017rtz, Raidal:2018bbj,Franciolini:2021tla,Liu:2021jnw,Franciolini:2022tfm,Escriva:2022bwe}) and detectable by future experiments \cite{Chen:2019irf,DeLuca:2021hde,Pujolas:2021yaw,Ng:2021sqn,Ng:2022agi,Ng:2022vbz}.

Primordial curvature perturbations comparable to those
necessary to generate a population of PBHs 
would also induce
GWs due to the nonlinear nature of gravity\,\cite{Tomita:1975kj,Matarrese:1993zf,Acquaviva:2002ud,Mollerach:2003nq,Ananda:2006af,Baumann:2007zm}.
The resulting stochastic GW background may be detected by GW observatories such as pulsar timing arrays (PTAs) \cite{NANOGrav:2020bcs,Goncharov:2021oub,Chen:2021rqp,Antoniadis:2022pcn}, the future Laser Interferometer Space Antenna \cite{LISACosmologyWorkingGroup:2022jok}, and ground-based observatories like LVK \cite{KAGRA:2013rdx} and the proposed 
Einstein Telescope and Cosmic Explorer \cite{Punturo:2010zz,Kalogera:2021bya,Maggiore:2019uih}. 

However, a model that features a sufficiently enhanced power spectrum at small scales to give rise to significant GWs and PBHs
would also lead to
the formation of an abundant population of highly dense dark matter minihalos \cite{Ricotti:2009bs,Kohri:2014lza,Gosenca:2017ybi,Delos:2017thv,Delos:2018ueo,Nakama:2019htb,Hertzberg:2019exb,Ando:2022tpj}, as long \chg{the dark matter is collisionless and capable of clustering}
on the relevant scales (e.g.~\cite{Bringmann:2009vf}).
While PBHs arise from $\mathcal{O}(1)$ initial density perturbations, perturbations as small as $\mathcal{O}(10^{-2})$ would still be sufficient to form halos already
before matter-radiation equality \cite{Kolb:1994fi,2010PhRvD..81j3529B,Berezinsky:2013fxa,Blanco:2019eij,StenDelos:2022jld}, which occurred at redshift $z\simeq 3400$ \cite{Planck:2018vyg}.
The smaller perturbations from which these halos arise would be far more common than the extreme perturbations necessary to produce PBHs, as illustrated in Fig.~\ref{fig:illustrate}.
Since these halos form
long before the redshift 30 to 50 at which halos would begin to form in a standard cold dark matter scenario (e.g.~\cite{Delos:2022bhp}),
they
would be characterized by extraordinarily high internal density. These ultradense halos would be extended compact objects that can survive to the present day, giving rise to new observational signatures.

Lensing observations have already been used extensively to search for evidence of PBHs by directly looking for the signatures of those compact objects on the magnification of observed stars (e.g. \cite{Paczynski:1985jf,EROS-2:2006ryy,Niikura:2017zjd,Niikura:2017zjd,
Katz:2018zrn,Niikura:2019kqi,
Montero-Camacho:2019jte,Blaineau:2022nhy,
Oguri:2022fir,Cai:2022kbp}). 
In this work, we address a different question: can we constrain scenarios with boosted small-scale power
(whether they produce PBHs or not)
by searching for the lensing signatures of the ultradense compact halos formed by the enhanced perturbations?\footnote{References~\cite{Dokuchaev:2002fm,Ricotti:2009bs} previously considered microlensing by halos arising in similar scenarios. Related approaches have also been proposed, including astrometric photolensing \cite{Li:2012qha} and distortions in strongly lensed images \cite{2013MNRAS.431.2172Z,Dai:2019lud}.}
Borrowing the analytic description of ultradense halo formation developed in Ref.~\cite{StenDelos:2022jld},  
we will show that large primordial curvature perturbations at 
$\mathcal{O}(0.1)$~pc scales, which correspond
to the formation of solar mass PBHs and nanohertz stochastic GW backgrounds,
can lead to observable lensing signatures.

\section{Ultradense dark matter halos from an enhanced curvature spectrum}\label{sec:formation}

\begin{figure}[t] 
    \centering
    \includegraphics[width=\columnwidth]{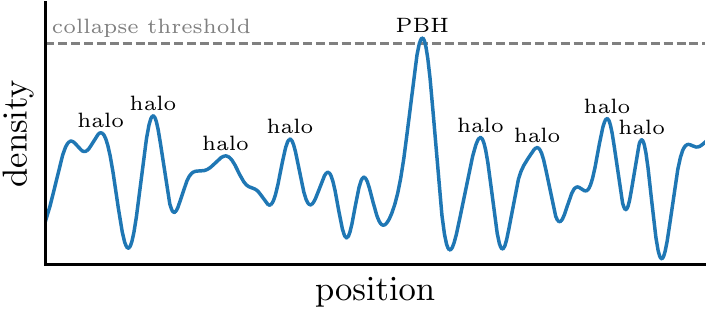}%
    \caption{Illustration of the ultradense halo scenario. We show a cartoon picture of the primordial density field. A region of extreme excess density is necessary to exceed the collapse threshold and produce a PBH (e.g.~\cite{Musco:2020jjb}), but more modest density excesses are much more common and form ultradense halos. Note that ultradense halos are not significantly correlated with PBHs spatially; they simply arise from the same cosmological scenario. Also, scenarios with too little primordial power to produce a significant number of PBHs can still yield abundant ultradense halos, as we will see in Sec.~\ref{sec:general}.}
    \label{fig:illustrate}
\end{figure}

In this section, we describe the abundance and properties of the ultradense dark matter halos formed in scenarios with an enhanced power spectrum at small scales. For concreteness, we consider the scenario described by model A of Ref.~\cite{Franciolini:2022tfm}, which produces PBHs around 10~M$_\odot$ comprising roughly $\approx 0.05\%$ of the dark matter and maximize the current upper bound set by LVK observations. 
Again, the ultradense halos are not connected to PBHs directly; they only emerge from the same cosmological scenario (see Fig.~\ref{fig:illustrate}). Ultradense halos can arise with or without PBHs, and
we will discuss implications for alternative (and agnostic) scenarios in Sec.~\ref{sec:general}.

Figure~\ref{fig:pk} shows the primordial curvature power spectrum $\mathcal{P}_\zeta(k)$ (dashed curve) in this model, which grows as $\mathcal{P}_\zeta\propto k^4$ at scales smaller than those constrained by CMB and large-scale structure data until it reaches a peak around $\mathcal{P}_\zeta\sim 10^{-2}$ near the pc$^{-1}$ scale. 
The PBHs form around that scale. We also show the matter power spectrum $\mathcal{P}(k,a)$ (solid curve) at $a=10^{-5}$, approximated as
\begin{equation}\label{eq:pk}
    \mathcal{P}(k,a) = I_1^2 \left[\log\left(\sqrt{2}I_2\frac{k}{\keq}\frac{a}{\aeq}\right)\right]^2 \mathcal{P}_\zeta(k) 
\end{equation}
with $I_1\simeq 6.4$ and $I_2\simeq 0.47$ \citep{Hu:1995en}. Here $\aeq\simeq 3\times 10^{-4}$ and $\keq\simeq 0.01$~Mpc$^{-1}$ are the scale factor and horizon wave number at matter-radiation equality, respectively. We assume for simplicity that dark matter is infinitely cold. In practice, for many dark matter models, the matter power should be truncated at some small scale due to kinetic coupling with the radiation and the resulting thermal free streaming (e.g.~\cite{Bertschinger:2006nq}). Our approach is thus limited to dark matter models that are capable of clustering at the relevant scales of about a comoving parsec.

\subsection{Ultradense halo formation}\label{sec:formationA}

\begin{figure}[t] 
    \centering
    \includegraphics[width=\columnwidth]{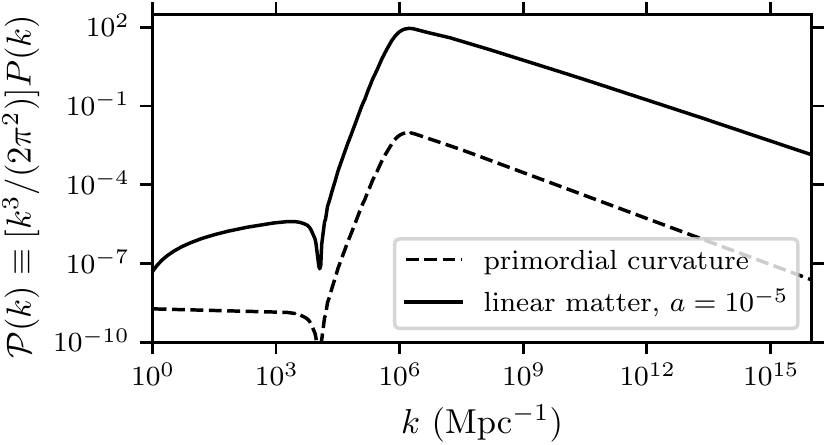}%
    \caption{Dimensionless power spectrum of primordial curvature perturbations (dashed curve) and matter perturbations (solid curve) in model A of Ref.~\cite{Franciolini:2022tfm}. The matter power spectrum is evaluated at $a=10^{-5}$, and the dark matter is assumed to be infinitely cold. Note that matter perturbations are already deeply nonlinear by this time.}
    \label{fig:pk}
\end{figure}

As Fig.~\ref{fig:pk} shows, the linear matter power spectrum $\mathcal{P}(k,a)\sim 10^2$ at $k\simeq 10^6$~Mpc$^{-1}$ by $a=10^{-5}$, implying that matter perturbations are already deeply nonlinear. During the radiation epoch, initially overdense regions exert gravitational attraction as they enter the horizon, which subsequently ceases as the radiation becomes homogeneous. However, dark matter particles set in motion by the initial pull continue drifting toward the initially overdense region, boosting its density further. In this scenario, Ref.~\cite{StenDelos:2022jld} showed that the linear density contrast threshold for collapse is
\begin{equation}\label{eq:dc}
    \dc = 3(1 + \sigma/\sqrt{5}),
\end{equation}
where $\sigma$ is the rms density contrast, if we approximate that the ellipticity $e$ of the initial tidal field within each region is equal to its most probable value,
\begin{equation}\label{eq:e}
    e=(\sqrt{5}\dc/\sigma)^{-1}
\end{equation}
(e.g.~\cite{Sheth:1999su}). As derived in Ref.~\cite{StenDelos:2022jld}, the collapse threshold in Eq.~(\ref{eq:dc}) leads to the excursion set mass function (e.g.~\cite{1991ApJ...379..440B})
\begin{equation}\label{eq:massfunction}
    \frac{\diff f}{\diff\log M}=\sqrt{\frac{2}{\pi}}
    \frac{(\nu+0.556)\e^{-\frac{1}{2}(\nu+1.34)^2}}{\left(1+0.0225\nu^{-2}\right)^{0.15}}
    \left|\frac{\diff\log\sigma_M}{\diff\log M}\right|
\end{equation}
describing the differential dark matter mass fraction in collapsed regions of mass $M$, where $\nu\equiv 3/\sigma_M$. Here $\sigma_M$ is the rms density contrast in spheres of mass $M$, i.e.
\begin{equation}\label{eq:sigma}
    \sigma_M^2=\int_0^\infty\frac{\d k}{k}\mathcal{P}(k)W^2(kr),
\end{equation}
with $W(x)\equiv 3(\sin x-x\cos x)/x^{3}$ and $M=(4\pi/3)\rhoMo r^3$, where $\rhoMo\simeq 33$~M$_\odot$\,kpc$^{-3}$ is the comoving dark matter density.\footnote{Since the power spectrum in Fig.~\ref{fig:pk} does not have a small-scale truncation, it is not necessary to employ the sharp $k$-space filtering used in Ref.~\cite{StenDelos:2022jld} (see Ref.~\cite{2013MNRAS.428.1774B}).}

\begin{figure}[!t] 
    \centering
    \includegraphics[width=\columnwidth]{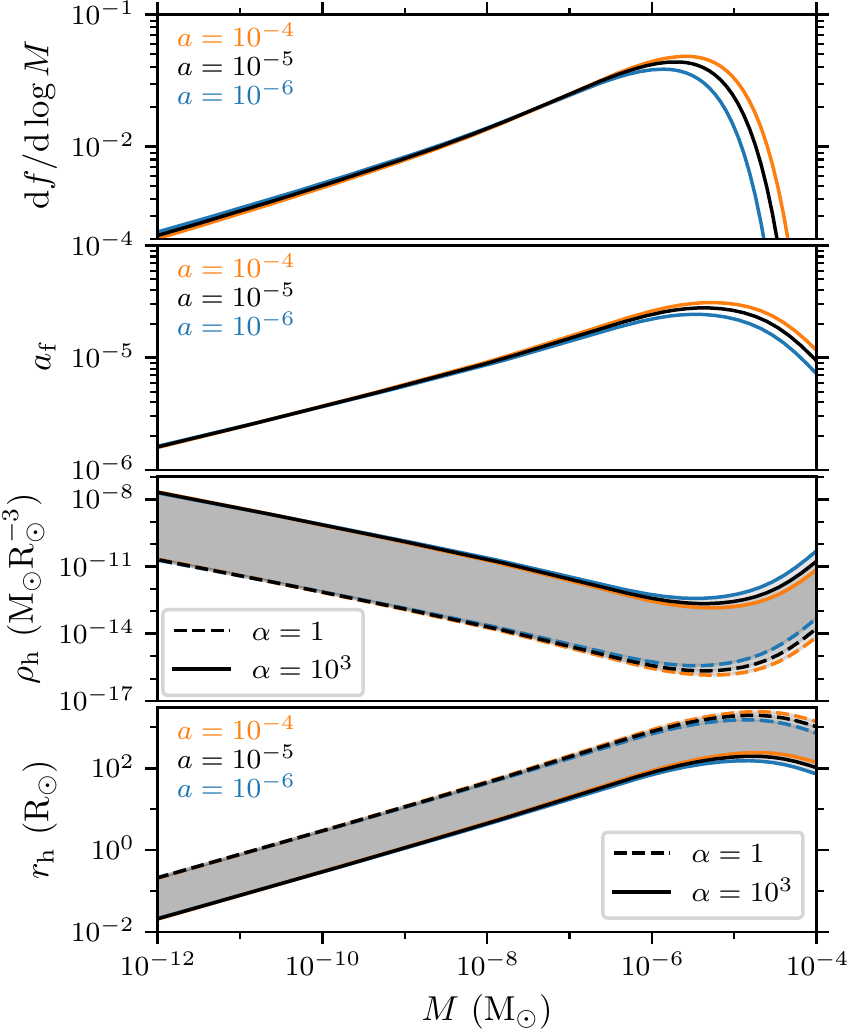}%
    \caption{Ultradense halos arising in a scenario characterized by the primordial power spectrum ${\cal P}_\zeta$ in Fig.~\ref{fig:pk}, at three different times during radiation domination.
    \textit{\textbf{Top panel:}} The differential dark matter mass fraction in collapsed regions of mass $M$. These regions become halos once they are locally matter dominated.
    \textit{\textbf{Upper middle panel:}} The typical scale factor $\af$ at which a collapsed region of mass $M$ becomes matter dominated, resulting in halo formation. That $\af$ depends on $a$ is an artifact of our simplified computation, but we note that $\af$ and $\d f/\d\log M$ are mostly independent of $a$.
    \chg{This independence breaks down when $M\gtrsim 10^{-6}$~M$_\odot$, but $\af\sim 10^{-5}$ in that regime, so for our analysis, we simply} adopt the values of $\af$ and $\d f/\d\log M$ at $a=10^{-5}$ (black curves).
    \textit{\textbf{Lower middle panel:}} Halo's characteristic density $\rhos$, derived from $\af$ as described in the text, as a function of mass $M$. We show both the conservative (dashed) and optimistic (solid) estimates; see Eq.~(\ref{eq:rho}).
    \textit{\textbf{Bottom panel:}} Likewise, the conservative (dashed) and optimistic (solid) estimates of a halo's scale radius $\rs$.}
    \label{fig:halos}
\end{figure}

Insofar as the collapse of matter density perturbations results in halo formation, $\d f/\d\log M$ describes the ultradense halo mass distribution. We evaluate $\d f/\d\log M$ for the power spectrum in Fig.~\ref{fig:pk} and plot it in the upper panel of Fig.~\ref{fig:halos}.
\chg{This distribution integrates to 28 percent of the dark matter at $a=10^{-5}$ (black curve), a much greater contribution than the sub-$10^{-4}$ fraction in PBHs. The peak of the distribution is}
around $\sim 10^{-6}$~M$_\odot$, a mass scale that differs from that of the  $\mathcal{O}(10)$~M$_\odot$ PBHs that emerge from the same scenario. When dark matter halos and PBHs arise from perturbations of the same scale, the PBH masses $M_\mathrm{PBH}$ are larger than the halo masses by a factor of about $\sim (\Meq/M_\mathrm{PBH})^{1/2}$ \cite{StenDelos:2022jld}, where $\Meq\simeq 3\times 10^{17}$~M$_\odot$ is the horizon mass at matter-radiation equality. This scaling arises because while halos form out of matter, PBHs form primarily out of radiation, the density of which far exceeds that of matter during the radiation epoch. 
This relation suggests that dark matter halos accompanying $\mathcal{O}(10)$~M$_\odot$ PBHs should have masses around $\sim10^{-7}$~M$_\odot$. In practice, halo masses somewhat exceed the above expectation because the halo formation threshold is much lower than that for PBHs, which allows halos to form from larger-scale initial perturbations.

However, 
the mass distribution $\d f/\d\log M$ derived above really describes the distribution of collapsed regions, and it cannot be taken for granted that these regions form halos.
During the radiation epoch, bound, virialized halos can only form within regions that are locally matter dominated \cite{Blanco:2019eij}. Reference~\cite{StenDelos:2022jld} showed that in the continued absence of gravitational forces, an initially overdense region achieves an 
overdensity $\rho/\rhoM$, 
with $\rhoM$ being the average matter density, as high as
\begin{equation}\label{eq:rhodrift}
 \frac{\rho}{\rhoM}
 \sim 
 e^{-2} 
 \left|
 \frac{\d\log\Xi}{\d\log r}
 \right|^{-1},   
\end{equation}
where $e$ is the ellipticity of the initial tidal field and 
\begin{equation}
\Xi(r)=3r^{-3}\int_0^r\xi(r^\p)r^{\p 2}\d r^\p    ,
\end{equation} 
with $\xi(r)$ being the correlation function. The functional dependencies in Eq.~(\ref{eq:rhodrift}) arise because a collapsing mass shell will simply overshoot and expand back outward if it does not produce a high enough matter overdensity to halt the expansion.
The following trends can be highlighted.
\begin{enumerate}[label={(\arabic*)},leftmargin=*]
\item 
The tidal field ellipticity $e$ enters because, without gravity, particle drifts must be highly focused to produce regions of significant overdensity.
\item 
The correlation function $\xi(r)$ enters because it describes the average radial profile of the density contrast about an arbitrary point. $\Xi(r)$ is then the average profile of the mean density enclosed within $r$. If $\Xi$ drops steeply (high $|\d\log\Xi/\d\log r|$), then the collapse of successively larger mass shells is gradual, and so the density contributed by later shells is not able to efficiently build on top of the density contributed by earlier shells before the early shells disperse away (after overshooting the collapse). If instead $\Xi$ drops shallowly (low $|\d\log\Xi/\d\log r|$), then the collapse of successively larger infalling mass shells occurs rapidly, and the density contributed by these shells builds up efficiently.
\end{enumerate}
Due to Eq.~(\ref{eq:rhodrift}), a collapsed region of mass $M$ becomes locally matter dominated, resulting in a halo formation, at roughly the scale factor
\begin{equation}\label{eq:af}
    \af(M) \sim e^{2} \left|\frac{r\int_0^\infty\d k\mathcal{P}(k)W^\p(k r)}{\int_0^\infty\frac{\d k}{k}\mathcal{P}(k)W(k r)}\right|\aeq,
\end{equation}
where $M=(4\pi/3)\rhoMo r^3$ again and $W^\p$ is the derivative of $W$. The fraction in Eq.~(\ref{eq:af}) is just $\d\log\Xi/\d\log r$. Meanwhile, Eqs. (\ref{eq:dc}) and~(\ref{eq:e}) imply that the ellipticity of the initial tidal field for a region of mass $M$, at its collapse time, is typically
\begin{equation}\label{eq:eM}
    e(M) = \frac{1}{3}\left[1 - \left( 1 + \frac{\sigma_M}{\sqrt{5}} \right)^{-1} \right].
\end{equation}
We show $\af(M)$, evaluated using Eqs. (\ref{eq:af}) and~(\ref{eq:eM}), in the second panel of Fig.~\ref{fig:halos}.

Note that, as formulated, $\af$ depends on the scale factor $a$ at which we evaluate the matter power spectrum $\mathcal{P}(k)$. This dependency is not physically meaningful, and in principle, $\af$ should instead depend on the full history of $\mathcal{P}(k)$. 
However, we will adopt the values of $\af$ and $\d f/\d\log M$ at $a=10^{-5}$ (black curves in Fig.~\ref{fig:halos}) for simplicity.
Both $\af$ and the halo mass function $\d f/\d\log M$ only vary significantly with $a$ 
\chg{for masses near $10^{-6}$~M$_\odot$, and since $\af\sim 10^{-5}$ there, the halo distributions evaluated at $a=10^{-5}$ are expected to be approximately correct.}

We also remark that the scaling of $\af$ with halo mass $M$ is easy to understand. In regimes where halos are rare, any halo must have formed from an extreme outlier peak in the initial density field. But outlier peaks are more spherical (e.g.~\cite{1986ApJ...304...15B}), which implies lower ellipticity $e$ and hence earlier halo formation according to Eq.~(\ref{eq:af}).

\subsection{Structures of ultradense halos}\label{sec:formationB}

\chg{We now discuss how we model the internal structures of the ultradense halos. Halo formation during the radiation epoch has not been simulated in detail, so there is considerable uncertainty in this treatment. Nevertheless, it is a general consequence of mass accretion in a cosmological context that the characteristic density $\rhos$ of a collisionless dark matter halo is closely linked to the mean density of the universe at its formation time $\af$ (e.g.~\cite{2010arXiv1010.2539D,2019PhRvD.100b3523D,Delos:2022yhn}). For halos that form during the radiation epoch, this motivates}
\begin{equation}\label{eq:rho}
    \rhos = \alpha \rhoRo \af^{-4},
\end{equation}
where $\rhoRo\simeq 0.012$~M$_\odot$\,kpc$^{-3}$ is the radiation density today and $\alpha$ is \chg{a} proportionality factor. Simulations during the matter epoch suggest 
\chg{that the density of material within a halo is about $10^3$ times the density of the universe at the time that the material became part of the halo \cite{2013MNRAS.432.1103L,Delos:2022yhn}.
This consideration suggests $\alpha\simeq 10^3$,}
but halo formation dynamics may be significantly different during the radiation epoch.
In the following, we will bracket the uncertainty in Eq.~(\ref{eq:rho}), as well as that in Eqs. (\ref{eq:rhodrift}) and~(\ref{eq:af}), by allowing $\alpha$ to vary.
Generally, one expects that at least the matter density does not drop during the formation process, which suggests $\alpha\sim 1$ as a lower limit. 
We will explore the consequences of
assuming $\alpha \in [1\divisionsymbol 10^3]$. The third panel of Fig.~\ref{fig:halos} shows $\rhos$ as a function of halo mass $M$ for $\alpha=10^3$ and $\alpha=1$.

We assume ultradense halos form with density profiles similar to the Navarro-Frenk-White (NFW) form \cite{1996ApJ...462..563N,1997ApJ...490..493N}
\chg{with scale radius $\rhos$ and scale density $\rs$.}
Since Fig.~\ref{fig:halos} shows halo mass functions evaluated close to their formation times, we set a halo's outer virial radius at this time to be $R_\mathrm{vir}=2\rs$, roughly the smallest value found in simulations of the smallest halos during the matter epoch \cite{2013JCAP...04..009A}. \chg{Integrating the NFW profile out to radius $2\rs$} leads to
\begin{equation}\label{eq:Mform}
    M\simeq 5.4\rhos\rs^3.
\end{equation}
Given $M$ and $\rhos$, this expression determines $\rs$. \chg{These choices are only intended to give an approximate representation of the inner structure of an ultradense halo, and their impact is secondary to the uncertainty in $\alpha$ in Eq.~(\ref{eq:rho}). For example,} the precise choice of $R_\mathrm{vir}/\rs$ has only a minor impact since the numerical coefficient in Eq.~(\ref{eq:Mform}) grows only logarithmically with the virial radius. The bottom panel of Fig.~\ref{fig:halos} shows the halo scale radii $\rs$ \chg{evaluated through this approach}. We also comment that $M$ here and in Fig.~\ref{fig:halos} thus represents the halo mass near the formation time and \emph{not} the mass today. Put in another way, $M$ is the mass of the densest central part of the halo, which is the part that contributes the most to microlensing and is the least susceptible to destruction during later evolution.

\begin{figure}[t] 
    \centering
    \includegraphics[width=\columnwidth]{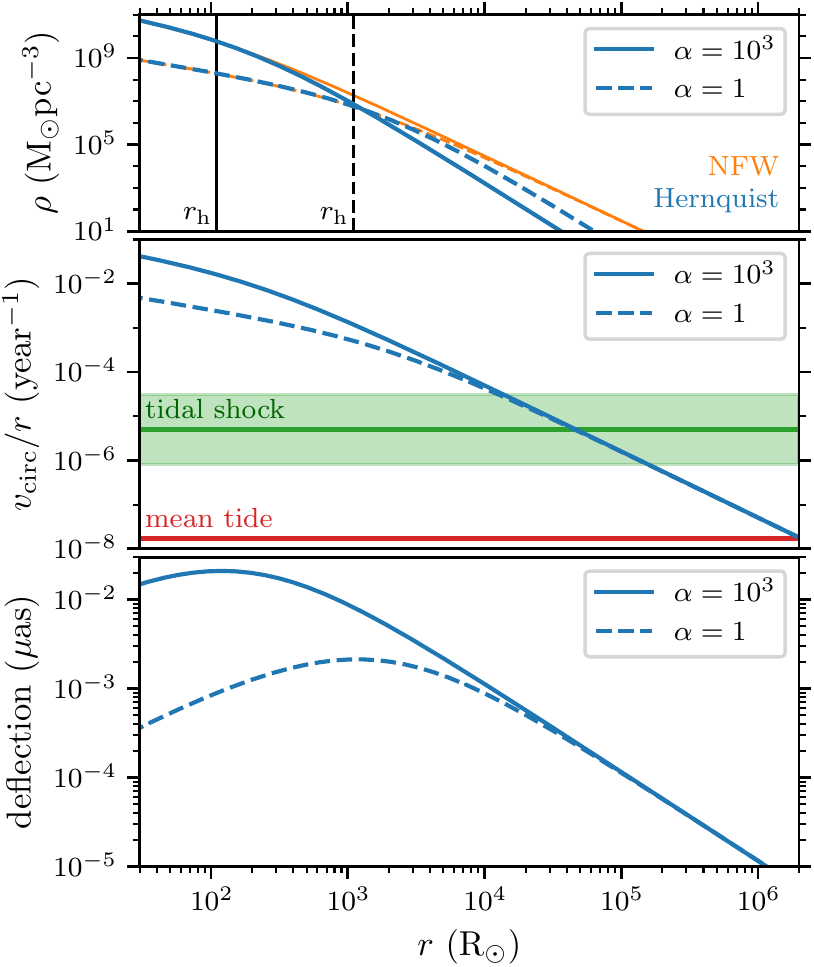}%
    \caption{Properties of an ultradense halo with formation mass $M=1.9\times 10^{-6}$~M$_\odot$ and characteristic radius $\rs=1100\alpha^{-1/3}$~R$_\odot$.
    \textit{\textbf{Upper panel:}}
    The density profile \chg{that we adopt [Hernquist form, Eq.~\eqref{eq:densprof}],} 
    in blue, for the optimistic and conservative assumptions of $\alpha=10^3$ (solid curves) and $\alpha=1$ (dashed curves), respectively. As Eq.~(\ref{eq:rho}) shows, $\alpha$ is the factor by which the halo's characteristic density exceeds that of the Universe at the halo's formation time. We also show in orange the corresponding NFW density profiles\chg{, for which the density is excessively high at large radii}.
    %
    \textit{\textbf{Middle panel:}} $v_\mathrm{circ}/r$ as a function of radius $r$ for the same density profiles, where $v_\mathrm{circ}$ is the circular orbit velocity. We show in green the median and central 68\% band for tidal shocking from stellar encounters for halos around 8~kpc from the Galactic Center, derived in Ref.~\cite{2023arXiv230104670S}. The interpretation is that where $v_\mathrm{circ}/r$ drops below the tidal shocking parameter, which is roughly the velocity kick $\Delta v/r$ imparted onto halo particles, disruption by stellar encounters is likely to become significant. We also show in red where the Galactic tidal forces become important.
    \textit{\textbf{Lower panel:}} The deflection angle for light passing the halo at closest distance $r$.
    }
    \label{fig:profile}
\end{figure}

\chg{While we assume that the NFW profile is approximately accurate near the halo's center, we cannot expect it to remain valid at distances much larger than $\rs$. The density profile in that regime is set long after halo formation by the details of the halo's accretion history \cite{2006MNRAS.368.1931L,2010arXiv1010.2539D,2013MNRAS.432.1103L,2019PhRvD.100b3523D}. It has been noted that}
the long-term accretion rates of galaxy cluster-scale halos \cite{2002ApJ...568...52W} are slow enough that $\rho\propto r^{-4}$ is predicted after sufficiently long times \cite{2006MNRAS.368.1931L}, as opposed to the NFW profile's $\rho\propto r^{-3}$. For ultradense halos in our scenario, the linear power spectrum has a similar $\mathcal{P}\propto k^{4}$ scaling \chg{to the spectrum of density variations relevant at cluster scales}, and we also anticipate that radiation domination during \chg{the early evolution of ultradense halos} will significantly suppress \chg{their} growth. Due to these considerations, we \chg{approximate the late-time} density profile of an ultradense halo with the Hernquist form \cite{1990ApJ...356..359H},
\begin{equation}\label{eq:densprof}
    \rho(r)
    = 
    \rhos(r/\rs)^{-1}(1+0.58 r/\rs)^{-3},
\end{equation}
where the numerical factor is tuned so that this profile closely matches the NFW profile
up to a few times $\rs$. At large radii, the profile in Eq.~(\ref{eq:densprof}) scales as $\rho\propto r^{-4}$ in accordance with the accretion rate considerations above. For a typical ultradense halo with mass $M=1.9\times 10^{-6}$~M$_\odot$ and characteristic radius $\rs=1100\alpha^{-1/3}$~R$_\odot$, the upper panel of Fig.~\ref{fig:profile} shows the density profile in Eq.~(\ref{eq:densprof}) as well as the corresponding NFW profile.

\chg{When integrated out to infinity, the mass associated with the density profile in Eq.~(\ref{eq:densprof}) converges to a value of about $3.5M$, where $M$ is the halo mass at formation given by Eq.~(\ref{eq:Mform}). We noted in Section~\ref{sec:formationA} that the ultradense halos contain about 28 percent of the dark matter at early times $a\sim 10^{-5}$. Thus, adoption of this density profile implies that ultradense halos eventually come to contain essentially all of the dark matter.
This outcome is consistent with reasonable expectations. For example, Ref.~\cite{2010MNRAS.401.1796A} found that in the standard picture of cosmological structure formation, 90 to 95 percent of the dark matter is expected to lie in halos today. A similar conclusion should apply to ultradense halos, which can accrete material up until around redshift 30, when CMB-level primordial perturbations (corresponding to $k\lesssim 10^4$~Mpc$^{-1}$ in Fig.~\ref{fig:pk}) finally become nonlinear and much larger cosmic structures begin to dominate. One limitation to our analysis, however, is that we assume every halo grows by the same factor of 3.5. A more detailed treatment would account for the spread in ultradense halo accretion histories.}

\chg{The central structures of collisionless dark matter halos remain mostly unaltered by later evolution (e.g.~\cite{Delos:2022yhn}). This is why the characteristics $\rhos$ and $\rs$ that we fix at the ultradense halos'} formation epoch are expected to remain largely accurate today. There are several ways in which this concordance can fail, however. First, ultradense halos can merge. While the merger rate is expected to be low for a power spectrum that grows as steeply as that of Fig.~\ref{fig:pk}, this effect would reduce the number of ultradense halos moderately. We note however that the fraction of dark matter in these halos is not altered, and simulations suggest that the characteristic density of a merger remnant is typically not lower than that of the progenitors \cite{2019MNRAS.487.1008D,2019PhRvD.100b3523D}. Thus, mergers are only expected to shift the ultradense halo distribution to slightly higher masses. We will neglect this effect here, leaving a more careful analysis for future work.

Ultradense halos also accrete onto much larger halos at later times, such as the halos that surround galaxies. Indeed, they are expected to contain a fraction of the dark matter inside galactic halos that is comparable to the fraction of dark matter that resides in ultradense halos initially. The extremely high density of these objects makes them essentially unaffected by tidal forces inside a much larger host. For example, the middle panel of Fig.~\ref{fig:profile} shows $v_\mathrm{circ}/r=\sqrt{F/r}$ as a function of radius $r$ for a typical ultradense halo, where $v_\mathrm{circ}$ is the circular orbit velocity and $F$ is the halo's central force. We compare this quantity to $\sqrt{\diff F_\mathrm{MW}/\diff R}$ (red line), where $\diff F_\mathrm{MW}/\diff R\simeq 300$~(km/s)$^2$/kpc$^2$ is the radial component of the Milky Way's tidal tensor at the Sun's galactocentric radius of 8~kpc, which we evaluate using the mass model in Ref.~\cite{2020MNRAS.494.4291C}. The interpretation is that the impact of Galactic tidal forces only becomes important when $v_\mathrm{circ}/r$ approaches $\sqrt{\diff F_\mathrm{MW}/\diff R}$, which occurs over $10^6$~R$_\odot$ from the halo center.

However, encounters with individual stars represent the more serious concern for ultradense halos inside the Galaxy. These become important when the velocity $\Delta v$ that they inject into halo particles approaches $v_\mathrm{circ}$. In the middle panel of Fig.~\ref{fig:profile}, we also show the distribution of tidal shocking parameters $B\sim \Delta v/r$, as defined and evaluated by Ref.~\cite{2023arXiv230104670S}, for halos orbiting the Galaxy at about 8~kpc. Shocks by stellar encounters are expected to significantly alter the structures of ultradense halos at radii beyond roughly $10^4$ to $10^5$~R$_\odot$. However, it is unclear what impact they have on $\rho\propto r^{-4}$ density profiles, because $\rho\propto r^{-4}$ is already the limiting density profile arising from such tidal shocks \cite{1987IAUS..127..511J,Delos:2021rqs}.\footnote{This behavior is a further theoretical motivation for adopting density profiles that scale as $\rho\propto r^{-4}$ at large radii in microlensing studies.} We will neglect stellar encounters in this work, with the justification that they are not expected to alter ultradense halo density profiles until radii significantly beyond $\rs$ are reached.\footnote{
Dark matter models characterized by 
a nonvanishing dark matter annihilation cross section would cause a depletion of the central, and densest, inner regions 
(see, e.g.,
\cite{Lacki:2010zf,Adamek:2019gns,Bertone:2019vsk,Carr:2020mqm,Kadota:2021jhg}).
However, this is not relevant to this work,
as the depletion typically takes place at $r\ll r\text{\tiny h}$.
}

Finally, the lower panel of Fig.~\ref{fig:profile} shows how our typical ultradense halo deflects passing light. We plot the deflection angle $(4G/c^2) M_\mathrm{2D}(r)/r$ as a function of the distance $r$ of closest approach, where
\begin{equation}\label{eq:M2D}
    M_\mathrm{2D}(r) \equiv
    \int_0^r 2\pi b\,\diff b 
    \int_{-\infty}^\infty\diff z\,
    \rho\!\left(\sqrt{b^2+z^2}\right)
\end{equation}
is the mass within an infinite cylinder of radius $r$.

\section{Microlensing constraints on ultradense halos}

In this section, we summarize the computation of the gravitational microlensing of light sources, applying the technique to constrain the ultradense minihalos described above. 
As dark matter halos are intrinsically extended objects, it is important to account for their size when deriving their potential lensing signatures. 
We follow the works of Refs.~\cite{Croon:2020ouk,Croon:2020wpr}
(see also \cite{Marfatia:2021twj,Fujikura:2021omw,Griest:1990vu,1994ApJ...430..505W,Fairbairn:2017sil})
to derive the rate of lensed events at the
EROS-2~\cite{EROS-2:2006ryy}, OGLE-IV~\cite{Niikura:2019kqi} and  Subaru-HSC~\cite{Niikura:2017zjd} surveys
while accounting for finite source sizes and extended lenses. 

\subsection{Detectability of a microlensing event}

\begin{figure}[t] 
    \centering
    \includegraphics[width=\columnwidth]{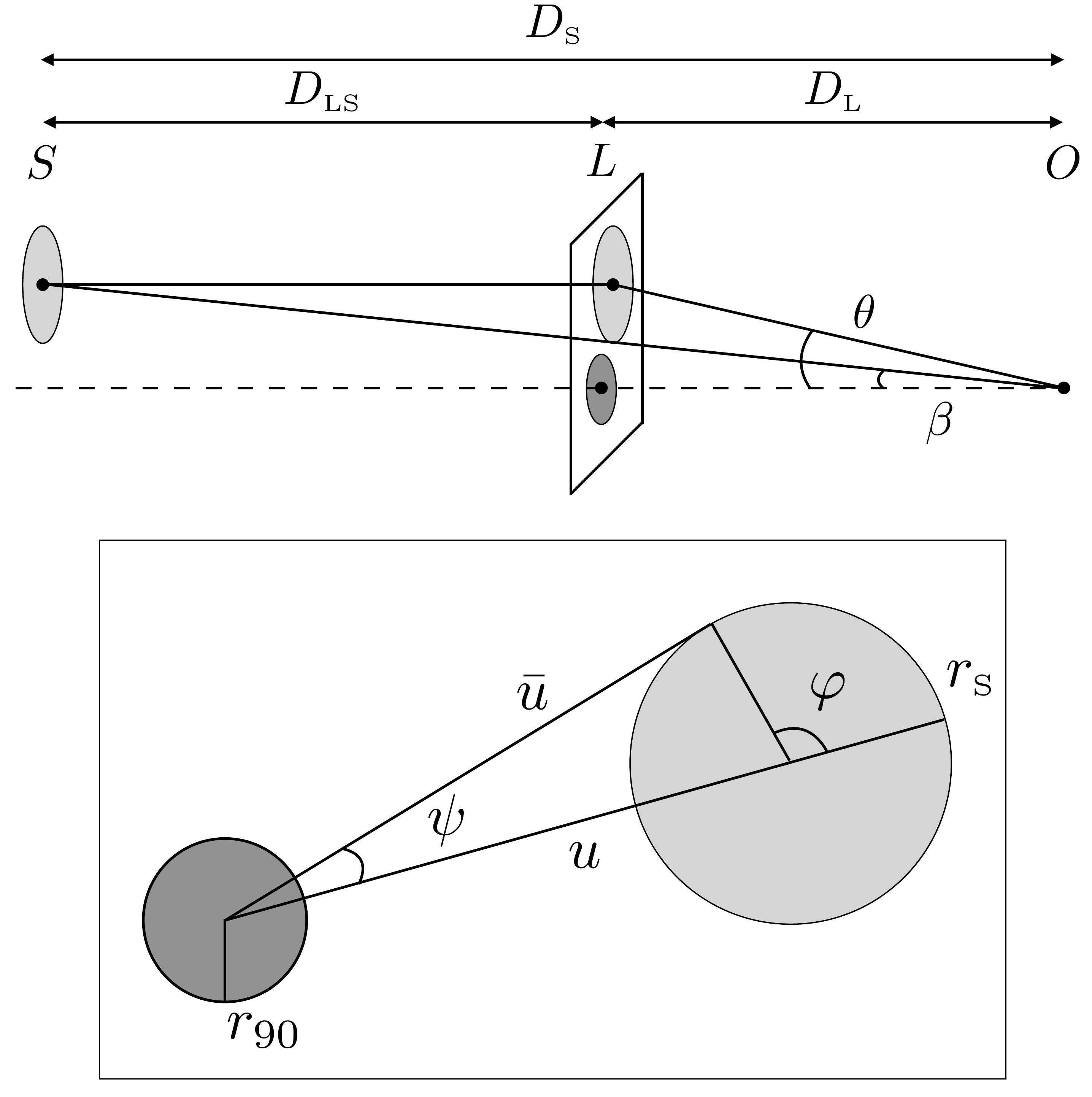}
    \caption{
\textit{\textbf{Upper panel:}}
Geometrical setup under consideration. The light source (S), the lens (L), and the observer (O) are separated by their respective distances $D_i$.
\textit{\textbf{Lower panel:}}
The lensing plane reporting the source and lens finite sizes, rescaled compared to the Einstein radius, as well as the integration angles. 
}
    \label{fig:setup_lensing}
\end{figure}

The light coming from a source is deflected by the gravitational field of an object (lens). 
For low-mass lenses, the deflection
cannot be resolved, but only a modification of the flux ${\cal F}$, defined as 
\begin{equation}
    \mu \equiv {{\cal F}}/{{\cal F}_0},
\end{equation}
may be detected, where ${\cal F}_0$ is the flux in the absence of lensing.
It is convenient to define the observer-lens, lens-source, and observer-source distances as $D_\text{\tiny L}$,
$D_\text{\tiny S}$, and $D_\text{\tiny LS} \equiv D_\text{\tiny S}-D_\text{\tiny L}$, respectively.
With respect to the axis passing through the lens center and the source, 
one can also define the angle $\beta$ as the true source position angle and $\theta$ as the angle of the observed lensed image of the source. We depict this geometrical setup in Fig.~\ref{fig:setup_lensing}.

As we noted above, the lensing halos are assumed to have Hernquist density profiles defined in Eq.~\eqref{eq:densprof}. 
These profiles have total mass $M_0=3.5M$, \chg{as we noted above,} where $M$ is the formation mass discussed in Sec.~\ref{sec:formation}.
Analogously, we define the late-time mass fraction, after halos have gained mass, as $f_0=3.5 f$. 
Also, we define 
$R_{90}=32\rs$ as the radius at which 90\% of the total mass is contained. These definitions mirror the choices used by Refs.~\cite{Croon:2020ouk,Croon:2020wpr}.

The lensing equation determines the path of light rays after a deflection and can be written as
\begin{equation}
    \beta=\theta-\frac{\theta_\text{\tiny E}^2}{\theta}\frac{M_0(\theta)}{M_0}~,
\label{eq:lens}
\end{equation}
where $M_0(\theta)=M_\mathrm{2D}(D_\text{\tiny L}\theta)$ is the lens mass projected onto the lens plane, as defined in Eq.~\eqref{eq:M2D}.
It is convenient to introduce the Einstein angle~\cite{Einstein:1936llh}
\begin{equation}
\theta_\text{\tiny E} 
\equiv\sqrt{\frac{4GM_0}{c^2}\frac{D_\text{\tiny LS}}{D_\text{\tiny L}D_\text{\tiny S}}}
=\sqrt{\frac{4GM_0}{c^2}
\frac{\left(1-x\right)}
{x D_\text{\tiny S}}
},
\label{eq:rE} 
\end{equation}
defined as the solution of the lensing equation when $\beta (\theta_\text{\tiny E})= 0$
and obtained as the value of $\theta$ for a pointlike lens $M_0(\theta) \rightarrow M_0$.
We also introduced the adimensional ratio $x \equiv D_\text{\tiny L}/D_\text{\tiny S}$.
Correspondingly, we denote the Einstein radius as the distance $\rE \equiv D_\text{\tiny L} \theta_\text{\tiny E}$ on the lens plane.
It takes values 
\begin{equation}
    R_\text{\tiny E}
    \simeq 2 \times 10^{3}
    R_\odot
    \lp \frac{M_0}{M_\odot}\rp^{1/2}
    \lp 
    \frac{D_\text{\tiny  S}}{10~{\rm kpc}}
    \rp^{1/2}
    \lp 
    \frac{1-x}{x}
    \rp^{1/2}.
    \label{eq:estRE}
\end{equation}

In units of $\rE$, the source radius in the lens plane is $r_\text{\tiny S}\equiv x R_\star/\rE$. In units of $\theta_\text{\tiny E}$,
the angular distance from the lens center to the source center is $u = \beta /\theta_\text{\tiny E}$
and to a point on the edge of the source is
\begin{equation}
\bar{u}(\varphi) =
\sqrt{u^2 + r_\text{\tiny S}^2 + 2 u r_\text{\tiny S} \cos\varphi}
\end{equation}
(see Fig.~\ref{fig:setup_lensing}). One can then rewrite the lensing Eq.~\eqref{eq:lens} for each infinitesimal point on the edge of the source as
\begin{equation}
\bar{u}(\varphi) = 
t - \frac{m(t)}{t},
 \label{eq:lensingeq_extednedlens}
\end{equation}
to find the positions of images at 
$t_i(\bar{u}(\varphi)) \equiv \theta_i/\theta_\text{\tiny E}$ 
with $i$ labeling the, in general, multiple solutions.
For a spherically symmetric density profile $\rho(r)$ we assume throughout this work, one can write
\begin{equation}
m(t) = \frac{\int_0^{t} d\sigma \sigma \int^\infty_0 d\lambda\, \rho(\rE\sqrt{\sigma^2+\lambda^2})}{\int_0^\infty d\gamma \gamma^2 \rho(\rE\gamma)}.
\label{eq:mt_def}    
\end{equation}

We neglect limb darkening and model the source star as having a uniform intensity in the lens plane.
It follows that the magnification produced by an image $i$ is given by the ratio of the image area to the source area \cite{1994ApJ...430..505W,Montero-Camacho:2019jte}
\begin{equation}
    \mu_i =  \frac{1}{4 \pi r_\text{\tiny s}^2}
    \llp 2 \eta 
   \int_0^{2\pi}
    {\rm d} \varphi \frac{\d\psi }{\d\varphi }
    \, t^2_i (\varphi)
    \rrp ,
\label{eq:magfinitesource}
\end{equation}
where
$\eta$ = sgn$(dt^2_i /d\bar{u}^2|_{\varphi = \pi})$ while the angular measure is defined from  the angle $\psi$ as
\begin{equation}
    \tan \psi \equiv \frac{r_\text{\tiny S}  \sin \varphi}{u+ r_\text{\tiny S} \cos \varphi}.
\end{equation}
Finally, we can compute the overall magnification $\mu_{\rm tot}$ 
as the sum of the individual contributions
\begin{equation}
    \mu_\text{\tiny tot} = \sum_i\mu_i.
\end{equation}
In this treatment, following Refs.~\cite{Smyth:2019whb,Croon:2020ouk},
we ignore the wave optics effects that are relevant when computing the magnification from lenses whose size is smaller than the wavelength of the detected light. 
For the masses considered in this work, the finite source size effect dominates the suppression of lensing signatures below $M \approx 10^{-11} M_\odot$ \cite{Sugiyama:2019dgt}.
Therefore, wave effects can be neglected.

If one takes the limit of negligible source size (i.e. $r_\text{\tiny S} \ll u$) and pointlike lens (i.e. $R_{90} \ll R_\text{\tiny E}$),
one can derive analytical solutions to the lens equation, and find
\begin{equation}
    \mu_\text{\tiny tot}  =
    \frac{ 2+u^2}{u\sqrt{u^2+4}}.
    \label{mupoint}
\end{equation}
In the opposite limit of a very large source $r_s\gg u$, one finds that the lensing solutions give a large suppression of $\mu$. This is because the lens only affects a negligible fraction of light rays coming from the source. 

\begin{figure}[!t!] 
    \centering
    \includegraphics[width=\columnwidth]{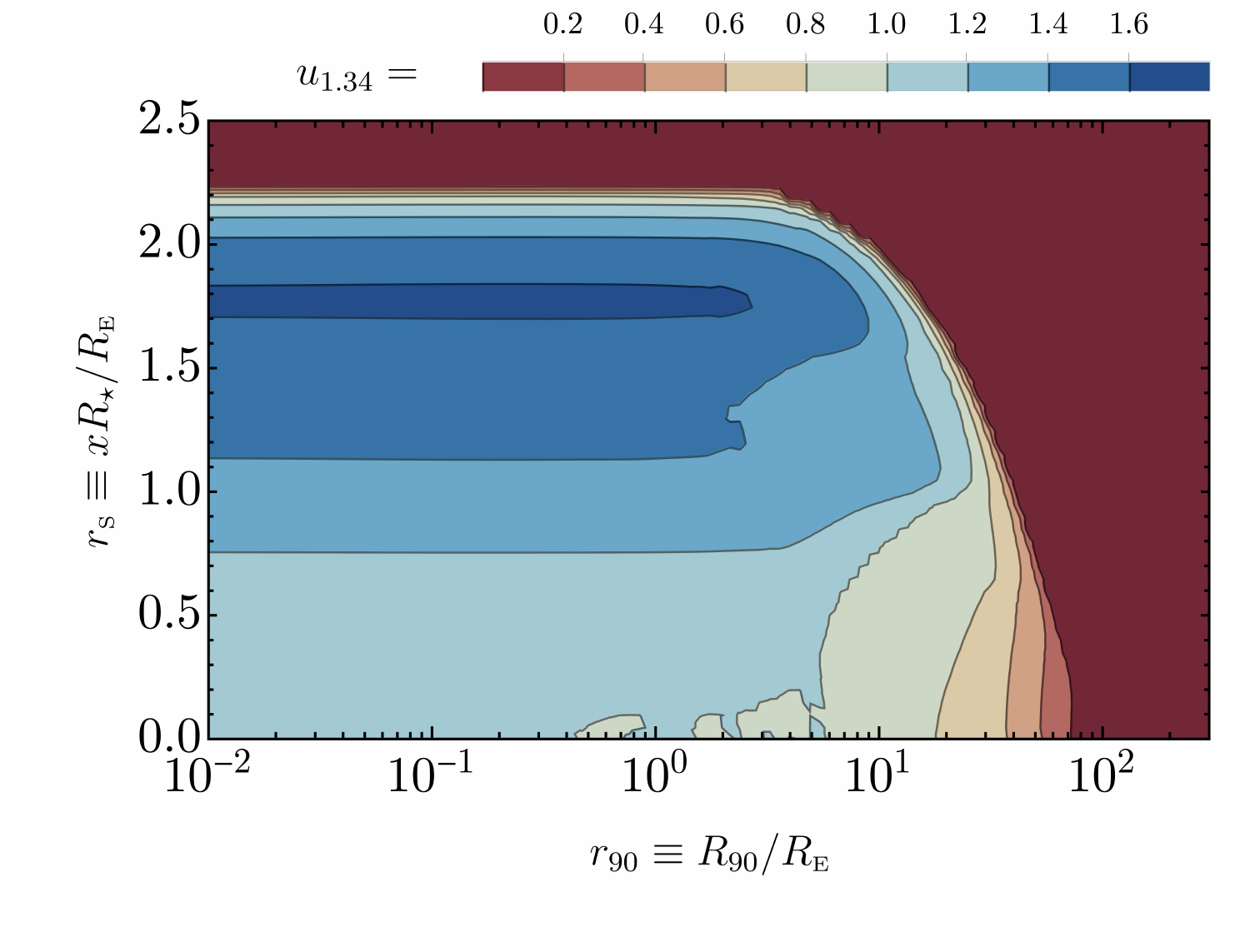}
    \caption{
    Threshold impact parameter for detection as a function of both the source size $r_\text{\tiny S}$ projected on the lens plane and the lens size $r_{90}$, both normalized with respect to the Einstein radius. 
    In the limit of negligible source and lens sizes, the threshold impact parameter converges toward the pointlike limit where $u_{1.34}\to 1$.
    }
    \label{fig:u134}
\end{figure}

Lensing surveys (such as EROS, OGLE, and HSC that we will consider later on)
define as detectable a microlensing event whose temporary magnification of the source star
exceeds the threshold value $\mu_\text{\tiny th}= 1.34$. 
Following this criterion, we will require 
$\mu_\text{\tiny tot}> \mu_\text{\tiny th}$. 
It is therefore convenient to generalize this criterion and  define the 
threshold impact parameter $u_{1.34}$
as
\begin{equation}
\mu_{\rm tot}(u\le u_{1.34}) \ge 1.34~,
\label{eq:uTdef}
\end{equation}
such that the magnification is above $1.34$ for all smaller impact parameters.
In the limit of a pointlike lens and negligible source size, one can directly derive from Eq.~\eqref{mupoint} the maximum impact parameter that satisfies this condition, which is $u = 1$.
We show in Fig.~\ref{fig:u134} the threshold impact parameter $u_{1.34}$ as a function of
both the source size $r_\text{\tiny S}$ projected on the lens plane and the lens size $r_{90}$.

\subsection{Number of detectable microlensing events}

The number of detectable lensing events can be computed by integrating the rate of overthreshold signals.
For a single source star and unit exposure time, 
the differential event rate with respect to the halo mass distribution, $x = D_\text{\tiny L}/D_\text{\tiny S}$,
and event timescale $t_\text{\tiny E}$ 
(i.e. the time the magnification remains
larger than the threshold), can be written as 
\begin{equation}
\frac{\d^2\Gamma}{\d x \d\tE \d \ln M_0} = 
\lp \frac{\d \rho_\text{\tiny lenses}(x)}{\d \log M_0} \rp 
\frac{2D_\text{\tiny S}\varepsilon(\tE) Q^2(x) }{M_0 v_0^2}
e^{-Q(x)/v^2_0}~,
\label{eq:dGammadxdt}
\end{equation}
where $v_0$ is the circular velocity in the galaxy. 
The differential density distribution of lenses $\rho_{\rm lenses}(x)$
can be derived by multiplying the galactic overdensity by the 
ultradense halo mass fraction, which means
\begin{equation}
 \frac{\d \rho_\text{\tiny lenses}(x)}{\d \log M_0} 
 \equiv 
\frac{\d f_0}{\d \log M_0}\times \rho_\text{\tiny DM}(x) .
\end{equation}
We also introduced 
$   Q(x) \equiv 4 \lp {u_{1.34}(x) \rE(x)}/{\tE}\rp^2$,
and $\varepsilon(\tE)$ is the efficiency of telescopic detection.
The total number of detectable events $N_\text{\tiny events}$ is 
\begin{align}
\frac{N_\text{\tiny events} }
{N_\star T_\text{\tiny obs}}
= 
\int 
{\rm d} \log M_0 
\d\Rstar 
\d\tE 
\d x
\lp
\frac{d^2\Gamma}{dx d\tE \d \log M_0}
\frac{\d n}{\d \Rstar}
\rp,
\label{eq:Nevents}
\end{align}
where $N_\star$ is the number of observed source stars in the survey, $T_\text{\tiny obs}$ is the total observation time, and $\d n/\d R_\star$ is the distribution of source star radii. 
As we will see, the finite source size is only relevant for the HSC survey of M31. 
When considering the other surveys, we will simply marginalize over the stellar radius distribution, as it does not affect the lensing signatures. 
In Appendix~\ref{app:Microsurv},
we summarise the setup of the three surveys we consider in this study, i.e.  
EROS-2~\cite{EROS-2:2006ryy}, OGLE-IV~\cite{Niikura:2019kqi} and  Subaru-HSC~\cite{Niikura:2017zjd}.

\begin{figure*}[t] 
    \centering
    \includegraphics[width=2\columnwidth]{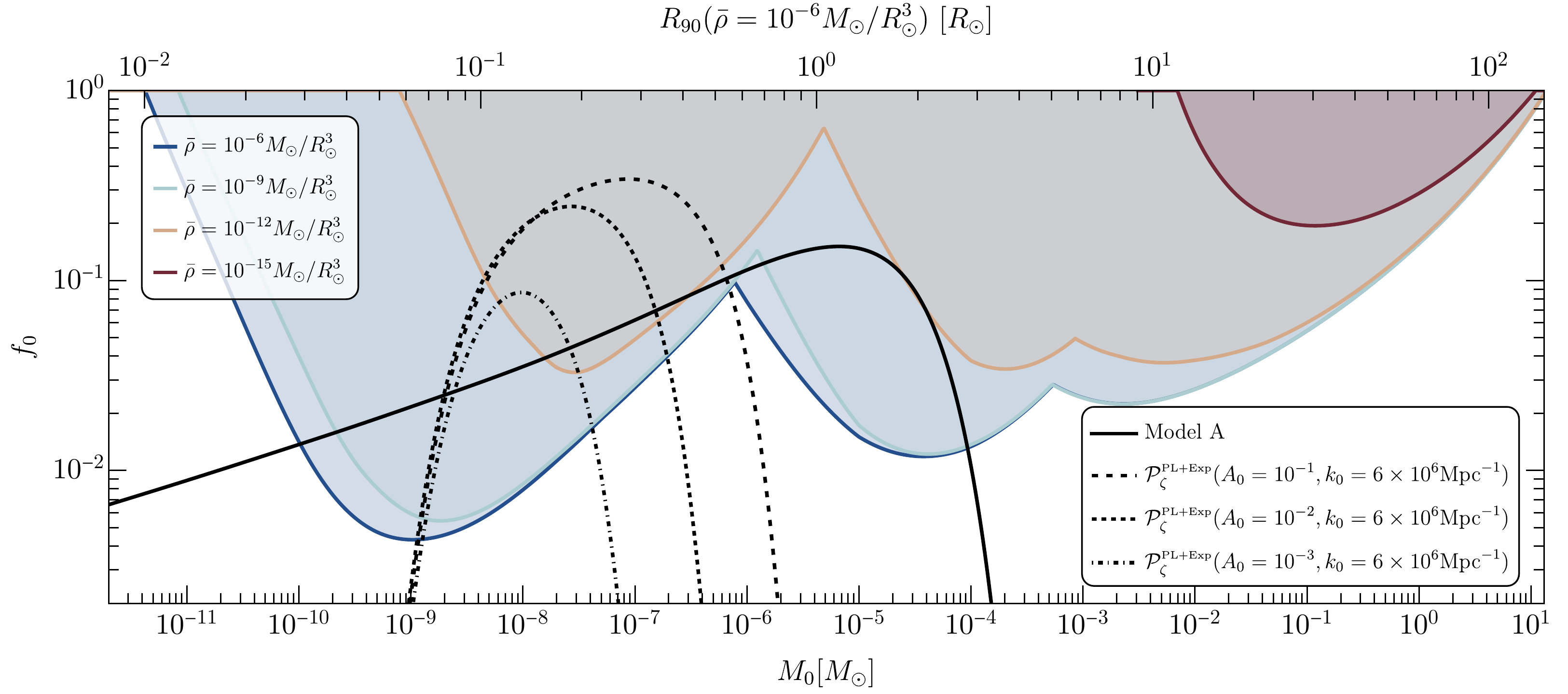}
    \caption{ 
    Upper bound on the fraction of dark matter in ultradense halos \chg{[with the density profile of Eq.~\eqref{eq:densprof}] for the simplified cases where all halos have the same mass and characteristic density}.
    The colored curves show the maximum mass fraction today, $f_0=3.5f$, assuming a monochromatic mass distribution at mass $M_0=3.5M$ and different values of the average halo density $\bar\rho=\rhos/7300$ ($f$, $M$, and $\rhos$ are the formation-time halo parameters considered in Sec.~\ref{sec:formation}).
    For each mass and density, the corresponding radius $R_{90}$ can be computed as $R_{90} = \lp 3M_0 /4 \pi\bar \rho\rp^{1/3}$. The upper side of the frame indicates the $R_{90}$ assuming $\bar \rho = 10^{-6} M_\odot /R_\odot^{3}$. 
    In this case, the halo sizes are always much smaller than the Einstein radius in Eq.~\eqref{eq:estRE}, and the constraints coincide with those derived in the point-mass limit (see Ref.~\cite{Croon:2020ouk}). 
    We also show, for visual comparison, the halo mass distribution $\d f_0 /\d \log M_0$ obtained in Sec.~\ref{sec:formation} considering model A from Ref.~\cite{Franciolini:2022tfm} as well as a few realizations of the narrowly peaked
    power spectrum in Eq.~\eqref{eq:pzeta_PLexp}.
    \chg{Note that when deriving constraints on these models, we will account for the full halo distributions arising therefrom rather than the monochromatic distributions probed here.}
    }
    \label{fig:constraint_mono}
\end{figure*}

To gain intuition on what is the reach of current constraints, in Fig.~\ref{fig:constraint_mono} we show the upper bound on the fraction of dark matter in the form of ultradense halos by taking a simplified monochromatic halo mass distribution.
We show the constraint by assuming different values of the average density $\bar \rho \equiv  3 M_0 / (4 \pi R_{90}^3) =\rhos/7300$ of the halos.
At the lowest masses, the dominant constraint comes from HSC data, peaking around $M_0 \simeq 10^{-9} M_\odot$. OGLE dominates the bounds at intermediate masses around $M_0 \simeq 5 \times 10^{-5} M_\odot$, 
while the EROS survey constrains the heavier portion of the plot. 
As one can see, for dense enough halos, 
the constraint converges to the one for pointlike lenses shown in Ref.~\cite{Croon:2020ouk}.
On the other hand, assuming less dense halos and correspondingly larger lens sizes relaxes the constraint, and the $M_0<10^{-2}M_\odot$ regime becomes entirely unconstrained for $\bar \rho \lesssim 10^{-15} M_\odot /R_\odot^{3}$.
This shows how crucial the halo density (or, equivalently, its physical extension) is for setting constraints via microlensing.

In Fig.~\ref{fig:constraint_mono} we superimpose the final halo mass distribution $\d f_0 /\d \log M_0$ obtained in Sec.~\ref{sec:formation} considering model A from Ref.~\cite{Franciolini:2022tfm}.
We see that this distribution crosses the constraint coming from HSC and OGLE  lensing surveys, but only if one assumes the average density to be at least as high as $\bar \rho = 10^{-12} M_\odot /R_\odot^{3}$.
However, translating the formation properties of the halos from Fig.~\ref{fig:halos} into the late-time Universe mass $M_0$ and size $R_{90}$, one discovers that halos in this scenario are too large to be constrained by the current experiment, having an average density of  $\bar \rho \approx 3 \times  10^{-17} M_\odot /R_\odot^{3}$ at $M_0 = 10^{-5} M_\odot$ and $\bar \rho \approx 7 \times  10^{-15} M_\odot /R_\odot^{3}$ at $M_0 = 10^{-8} M_\odot$, assuming $\alpha = 10^3$. Even smaller average densities, and correspondingly larger lens sizes, are attained by assuming smaller values of $\alpha$.
We conclude, therefore, that current lensing surveys are not able to constrain enhanced power spectra peaked around $k\approx 10^6/{\rm Mpc}$. 
In particular, \chg{we verified by considering the full halo distributions from Fig.~\ref{fig:halos} (as opposed to the monochromatic cases studied in Fig.~\ref{fig:constraint_mono}) that} the specific PBH formation scenario in the LVK detection window from Ref.~\cite{Franciolini:2022tfm} (model A) is not currently constrained by lensing surveys.

Also, due to the strong relationship between a halo's density and its mass in Fig.~\ref{fig:halos}, the strongest prospects for the detection of ultradense halos in such a PBH scenario involve using HSC to constrain the low-mass tail of the halo distribution. Unfortunately, since this tail is a highly model-dependent feature (related to how shallowly the primordial power spectrum in Fig.~\ref{fig:pk} decays at large $k$), this approach is unlikely to yield generally applicable constraints on the PBH scenario.
In the next section, we will explore the constraints that can be set by current and future HSC observations on a narrow enhancement to the power spectrum, for which this low-mass halo tail does not contribute.
 
\section{Constraints on the power spectrum at small scales}\label{sec:general}

We now test the sensitivity of microlensing to the ultradense halos arising from a broader family of primordial curvature power spectra. 
We want to consider realistic narrow spectra as a benchmark. Therefore, we consider
\begin{equation}
 \mathcal{P}_\zeta^\text{\tiny PL+Exp}(k) = 
 A_0 (k/k_0)^{4} \exp[2-2(k/k_0)^2],
 \label{eq:pzeta_PLexp}
\end{equation}
which is parametrized by the peak amplitude $A_0$ and wave number $k_0$ such that the maximum is achieved at $\mathcal{P}_\zeta^\text{\tiny PL+Exp}(k_0)=A_0$. This spectrum grows as $\mathcal{P}_\zeta(k)\propto k^4$ for $k<k_0$, the characteristic growing slope that can arise in simple ultraslow-roll inflation models (see e.g. \cite{Byrnes:2018txb,Franciolini:2022pav,Karam:2022nym}), while it is Gaussian suppressed for $k>k_0$.
The precise form of the small-scale (large $k$) suppression turns out to be unimportant for our constraint. We explicitly check this by also considering the functional form
\begin{equation}
    \mathcal{P}_\zeta^\text{\tiny PL+PL}(k) =
    {2A_0}/\llp {(k/k_0)^{-4}+(k/k_0)^4}\rrp^{-1}.
 \label{eq:pzeta_PLPL}
\end{equation}
This spectrum  similarly peaks at $\mathcal{P}_\zeta^\text{\tiny PL+PL}(k_0)=A_0$ and grows as $k^4$ for $k<k_0$, but at larger $k$ it decays as $k^{-4}$.

We repeat the procedure in Sec.~\ref{sec:formation} to generate the ultradense halo distribution for these scenarios. We make one change, however. Press-Schechter mass functions like Eq.~(\ref{eq:massfunction}), \chg{when evaluated using the real-space top-hat window function,} are not well behaved when the power spectrum decays rapidly at small scales. \chg{They predict a halo count that diverges at small mass scales, even when there is no power on such scales \cite{2013MNRAS.428.1774B}. This difficulty arises from the assumption of uncorrelated steps in the excursion set formulation of Press-Schechter theory \cite{1991ApJ...379..440B}, which corresponds to the use of a sharp $k$-space window function, $W(x)=\step(c-x)$, instead of the top-hat window. Here $c$ is a constant, which fixes the connection between the wavenumber and the mass scale, and $\step$ is the Heaviside step function. Therefore, we adopt the sharp $k$-space window when evaluating $\sigma_M$ in Eq.~(\ref{eq:sigma}). Following Refs.~\cite{1993MNRAS.262..627L,2013MNRAS.428.1774B}, we set $c=2.5$.}

\begin{figure}[!t] 
    \centering
    \includegraphics[width=\columnwidth]{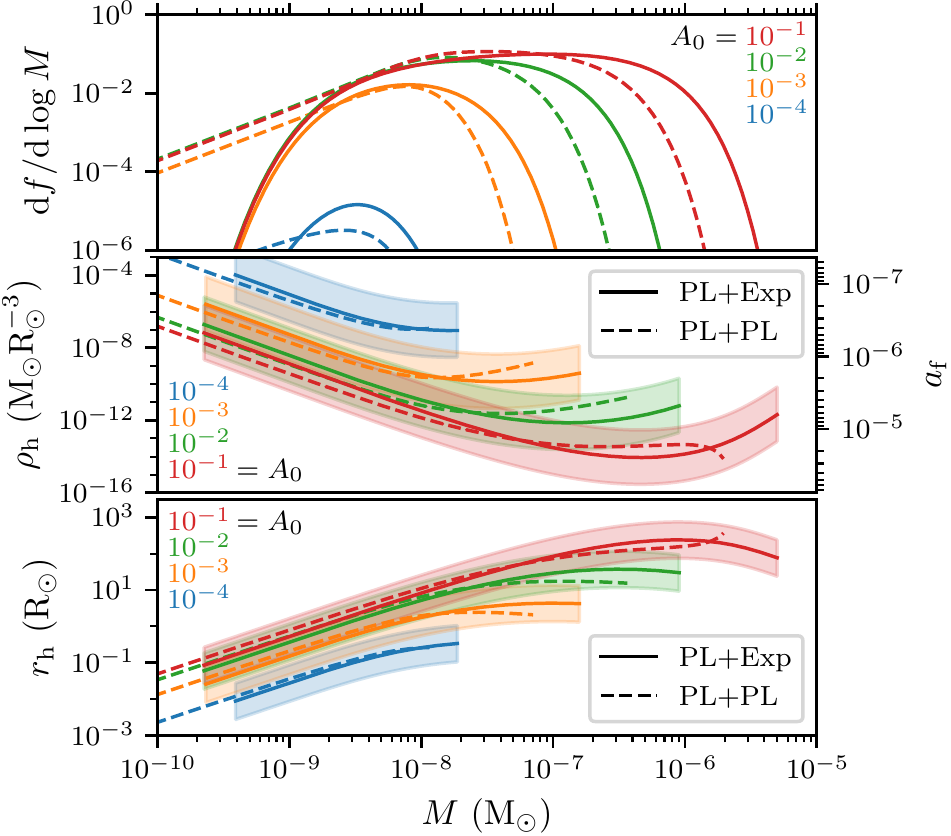}%
    \caption{Ultradense halos arising from the PL+Exp power spectrum [Eq.~\eqref{eq:pzeta_PLexp}, solid curves] and the PL+PL power spectrum [Eq.~\eqref{eq:pzeta_PLPL}, dashed curves]. We fix $k_0=6\times 10^{6}$ Mpc$^{-1}$ but vary the amplitude $A_0$ from $10^{-4}$ to $10^{-1}$ (different colors).
    \textit{\textbf{Upper panel:}} The differential dark matter mass fraction in halos as a function of formation mass $M$.
    \textit{\textbf{Middle panel:}} Halo's characteristic density $\rhos$, as a function of formation mass $M$, for $\alpha$ ranging from 1 to $10^3$ [shaded bands; see Eq.~\eqref{eq:rho}]. The lines correspond to $\alpha=30$; for the PL+PL power spectrum (dashed) we only show this line. The right-hand axis also denotes the halo formation time (the bands are not relevant here).
    \textit{\textbf{Lower panel:}} Likewise, a halo's scale radius $\rs$ as a function of formation mass $M$.}
    \label{fig:halos_narrow}
\end{figure}

Figure~\ref{fig:halos_narrow} shows the resulting halo distributions.
One noteworthy feature is that, even though the overall mass fraction $f$ 
decreases with smaller $A_0$,
the corresponding halo density $\rhos$ increases. 
This is a consequence of the important role of ellipticity in the collapse; see Sec.~\ref{sec:formation} for more details. 
As the amplitudes of primordial perturbations decrease, the overdensities that can collapse to form ultradense halos are rarer and hence increasingly spherical (e.g.~\cite{1986ApJ...304...15B}). This allows the halos to form at earlier epochs and therefore possess larger internal density.

We derive current constraints and forecast what is accessible with future HSC observations (see for example forecasts in Ref.~\cite{Kusenko:2020pcg}) by following the same steps as in the previous section. 
In particular, we assume $T_\text{\tiny obs} = 7$~h to derive the constraint using available HSC observations~\cite{Niikura:2017zjd} and  $T_\text{\tiny obs} = 70$~h to forecast future reach (extrapolating the same detection efficiency and assuming the same number of stars in the survey).

\begin{figure}[!h] 
    \centering
      \includegraphics[width=\columnwidth]{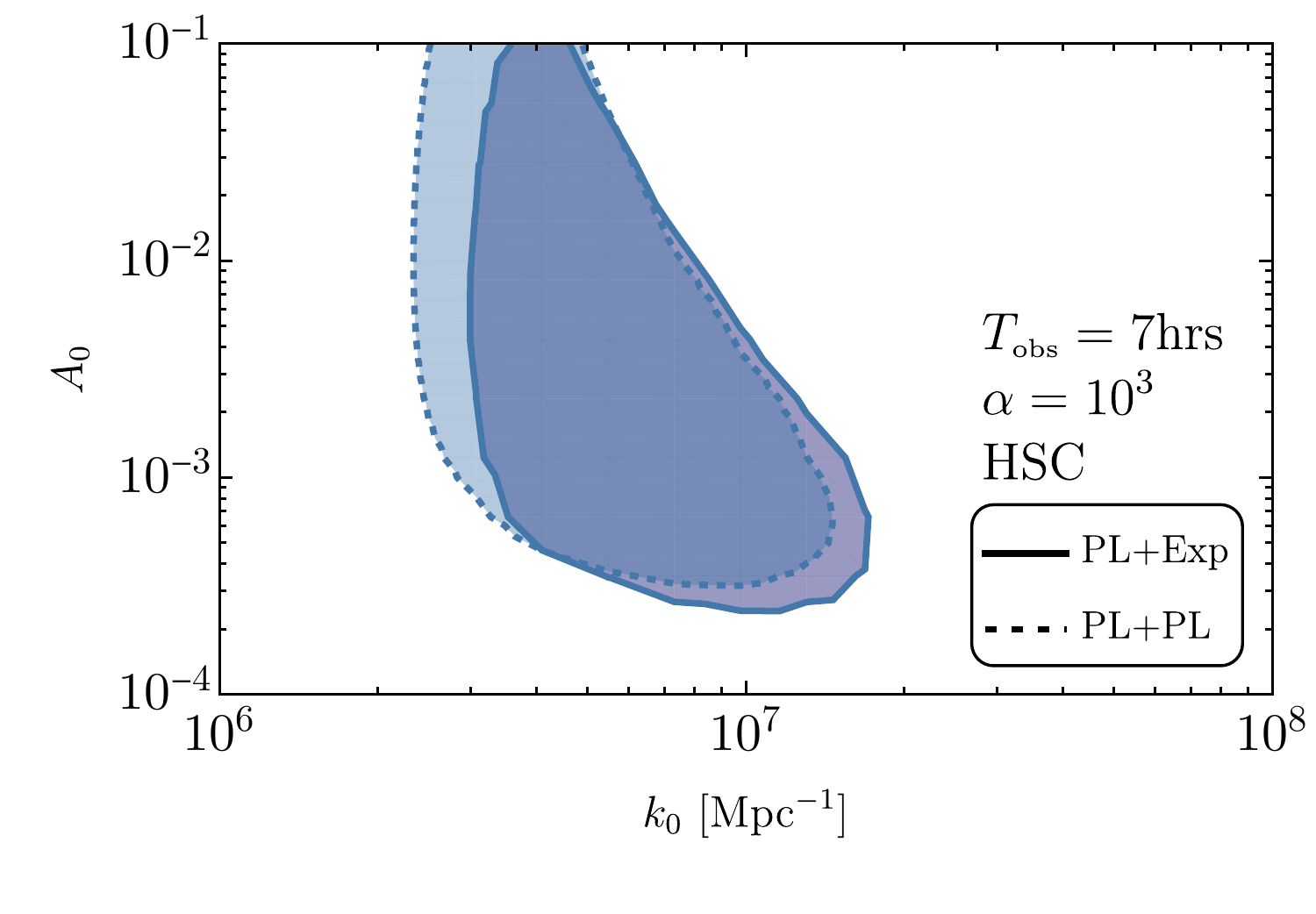}%
    \caption{
    Comparison between HSC constraints on the PL+Exp or PL+PL power spectra in Eqs.~\eqref{eq:pzeta_PLexp} and~\eqref{eq:pzeta_PLPL} that peak at amplitude $A_0$ at the scale wave number $k_0$. 
    We shade the region excluded by current HSC observations with $T_\text{\tiny obs} = 7$~h under the optimistic assumption $\alpha= 10^3$ about ultradense halos' internal structures. Evidently, the two parametrizations yield comparable results.
    }
    \label{fig:constraints_narrowPS_0}
\end{figure}

We first test the impact of different parametrizations of the spectral shape, Eqs. (\ref{eq:pzeta_PLexp}) and~(\ref{eq:pzeta_PLPL}). Figure~\ref{fig:constraints_narrowPS_0} shows the current HSC constraint for both cases under the optimistic assumption that $\alpha=10^3$ [see Eq.~\eqref{eq:rho}]. The close match between the two outcomes confirms that the particular form of the low-mass tail of the power spectrum is not important.

Figure~\ref{fig:constraints_narrowPS} shows the HSC constraints for the $\mathcal{P}_\zeta^\text{\tiny PL+Exp}(k)$ power spectrum parametrization [Eq.~\eqref{eq:pzeta_PLexp}].
Under the optimistic assumption that $\alpha=10^3$, current measurements force the spectral amplitude to be $A_0\lesssim 2\times  10^{-3}$ around $k_0 = 10^7$ (solid blue curve).
Increasing the observation time $T_\text{\tiny obs}$ by a factor of 10 would give rise to slightly more stringent constraints (solid red curve) in a wider range of $k_0$. 
We also note that for larger spectral amplitudes $A_0$, the constraint degrades. This is a consequence of the more diffuse halos that arise if the amplitudes of initial density perturbations are too large, as discussed above.

As we discussed in Sec.~\ref{sec:formation}, $\alpha$ parametrizes theoretical uncertainty about the internal structures of ultradense halos. If we adopt a more moderate assumption, $\alpha =30$, then the current constraint disappears and only future observations can constrain a smaller portion of parameter space (red dashed curve). With the most conservative assumption of $\alpha =1$ instead, both current and future constraints disappear, as the lenses are then too diffuse to generate observable signatures within the HSC survey. This outcome motivates further study of the ultradense halo formation scenario, likely with numerical simulations, to understand their internal structures and hence determine whether they can produce solid constraints through microlensing.

\subsection{Consequences for PBHs and GWs}

In this section, we briefly summarize the potential consequences of these constraints on scenarios of PBH formation as well as the production of induced GWs.

\subsubsection{Primordial black holes}

To compare lensing constraints with the PBH scenario, we compute the power spectral amplitude required to generate a significant abundance of PBHs assuming both spectra defined in Eqs.~\eqref{eq:pzeta_PLexp} and \eqref{eq:pzeta_PLPL}.

The first step is to consider the relationship
\begin{equation}
    M_H \simeq  
 17 M_\odot\left(\frac{g_*}{10.75}\right)^{-1/6}
 \left(\frac{k/\kappa}{{\rm pc}^{-1}}\right)^{-2}
 \label{M-k}
\end{equation}
between the cosmological horizon mass $M_H$, which relates to the PBH mass $m_\text{\tiny PBH}$ by an order unity factor dictated by the critical collapse parameters \cite{Escriva:2021aeh},
and the comoving wave number $k$. Here, $g_*$ is the effective number of degrees of freedom of relativistic particles and $\kappa \equiv k r_m $ relates $k$ to the characteristic perturbation size at horizon crossing $r_m$.
For example, if we consider a spectrum centered around $k_0 = 6 \times 10^6~{\rm Mpc}^{-1}$
one finds $M_H \approx 2.5~M_\odot$, corresponding to the formation of PBHs of around a solar mass.

We compute the PBH abundance $f_\text{\tiny PBH}$ 
as \cite{Sasaki:2018dmp}
\begin{align}\label{eq:fPBH}
f_\text{\tiny PBH}\equiv 
\frac{\Omega_\text{\tiny PBH}}{\Omega_\text{\tiny CDM}} 
=
\frac{1}{\Omega_\text{\tiny CDM}}
\int \d \log M_H 
\left(
\frac{M_\text{\tiny eq}}{M_H}
\right)^{1/2}\beta(M_H),
\end{align}
where 
 $M_H$ is the horizon mass at the time of horizon reentry, 
 $M_\text{\tiny eq} \simeq 3 \times 10^{17}$ $M_{\odot}$ the horizon mass at matter-radiation equality,
and $\Omega_\text{\tiny  CDM}$ is the dark matter density today (in units of the critical density).
Adopting threshold statistics, we compute the mass fraction assuming Gaussian primordial curvature perturbations\footnote{
See, however, the recent Refs.~\cite{Ferrante:2022mui,Gow:2022jfb} for nonperturbative extensions of this computation if one assumes 
non-Gaussian primordial curvature perturbations. 
} and accounting for the nonlinear relationship between curvature and density perturbations \cite{DeLuca:2019qsy,Young:2019yug,Germani:2019zez}.
One obtains
\begin{align}
\beta(M_H) & 
= 
\mathcal{K}
\int_{\delta_l^{\rm min}}^{\delta_l^{\rm max}}
\d \delta_l
\left(\delta_l - \frac{1}{4\Phi}\delta_l^2 - \delta_c\right)^{\gamma}
P_\text{\tiny G}(\delta_l),
\label{eq:GaussianTerm1}
\\
P_\text{\tiny G}(\delta_l) & = \frac{1}{\sqrt{2\pi}\sigma(r_m)}e^{-\delta_l^2/2\sigma^2(r_m)},
\label{eq:GaussianTerm}
\end{align}
where $\delta_l$ is the linear (i.e. Gaussian) component of the density contrast and the integration boundaries are dictated by having overthreshold perturbations and type-I PBH collapse (see e.g.~\cite{Musco:2018rwt}).
We indicate with $\sigma(r_m)$ the variance of the linear density field computed at horizon crossing time and smoothed on a scale $r_m$ (see e.g. Ref.~\cite{Franciolini:2022tfm} for more details), while $\delta_c$ is the threshold for collapse.
We also introduced the parameters ${\cal K}$ and $\gamma$ to include the effect of critical collapse, while $\Phi$ controls the relationship between the density contrast and the curvature perturbations.

We adopt the technique of Ref.~\cite{Musco:2020jjb} to compute the threshold $\delta_c$ for PBH formation.
The PL+Exp spectrum [Eq.~\eqref{eq:pzeta_PLexp}] gives rise to collapsing peaks for which the characteristic comoving size is
$\kappa \equiv  r_m k = 2.51 $, 
the shape parameter is $\alpha_c= 4.14$ (see Ref.~\cite{Musco:2018rwt} for more details) and the threshold for collapse is $\delta_c= 0.572$ in the limit of perfect radiation domination.
When considering the PL+PL spectrum [Eq.~\eqref{eq:pzeta_PLPL}], we find instead that 
$\kappa= 2.36 $,  $\alpha_c= 3.25$, and $\delta_c= 0.558$.

We include the effect of the softening of the equation of state of the Standard Model plasma due to the QCD transition as done in Ref.~\cite{Franciolini:2022tfm} and based on the numerical simulations of Ref.~\cite{Muscoinprep} (see also \cite{Byrnes:2018clq,Carr:2019kxo,Escriva:2022bwe}).
 This generates the slight dip in the black lines of Fig.~\ref{fig:constraints_narrowPS} around $M_H\approx M_\odot$.

The gray shaded region at the top of Fig.~\ref{fig:constraints_narrowPS} is immediately ruled out because PBHs would be produced with an abundance larger than all of the dark matter in our Universe ($f_\text{\tiny PBH}\geq 1$). The two lines below indicate PBH mass fractions $f_\text{\tiny PBH}= 10^{-3}$ and $f_\text{\tiny PBH}= 10^{-5}$ respectively, following the logarithmic scaling with $A_0$.  
We conclude from the plot that, provided that dark matter can cluster on the relevant scales and that halos are dense enough with $\alpha\approx 10^3$, current HSC data exclude the possibility that a narrow population of PBHs with stellar mass comprise a non-negligible fraction of the dark matter.

\subsubsection{Induced stochastic GW background}

We also compute the GW signal sourced by scalar perturbations at second order to show how constraints on ultradense dark matter halos may have interesting consequences for induced GWs within the nanohertz frequency range \cite{Vaskonen:2020lbd,DeLuca:2020agl,Kohri:2020qqd,Inomata:2020xad,Zhao:2022kvz}.
The frequency $f_\text{\tiny GW}$ of the stochastic GW background (SGWB) is related to the comoving wave number $k =  2 \pi f_\text{\tiny GW}$ by the relation 
\begin{align}
k \simeq 6.47\times 10^{14} \left(
\frac{f_\text{\tiny GW}}{{\rm Hz}}
\right)\,\,{\rm Mpc}^{-1}.
\end{align}
For example, a spectrum centered around $k_0 = 6 \times 10^6~{\rm Mpc}^{-1}$ corresponds to $f_\text{\tiny GW} \approx 9 \times 10^{-9}$~Hz.
This falls within the range of frequencies 
corresponding to the putative signal recently reported by the NANOGrav Collaboration~\cite{NANOGrav:2020bcs} (and also independently supported by other pulsar timing array data \cite{Goncharov:2021oub,Chen:2021rqp,Antoniadis:2022pcn}).

The current energy density of 
GWs is given by \cite{Tomita:1975kj,Matarrese:1993zf,Acquaviva:2002ud,Mollerach:2003nq,Ananda:2006af,Baumann:2007zm}
\begin{align}
 \Omega_{\text{\tiny GW},0} & = 0.39\, \Omega_{r,0}
 \left[\frac{g_*(T_H)}{106.75}\right]
 \left[\frac{g_{*,s}(T_H)}{106.75}\right]^{-\frac{4}{3}}
 \Omega_{\text{\tiny GW},H}
 \label{eq:cg}
 \end{align}
 as function of their frequency $f_\text{\tiny GW}$, with 
 \begin{align}
 & \Omega_{\text{\tiny GW},H} = \label{eq:Transfer}
 \left(\frac{k}{k_H}\right)^{-2b}
 \int_{0}^{\infty}\!\!\d v
 \int_{|1-v|}^{1+v}\!\!\d u
 \mathcal{T}(u,v)
 \mathcal{P}_{\zeta}(ku)\mathcal{P}_{\zeta}(kv)
\end{align}
(see e.g. Ref.~\cite{Domenech:2021ztg} for a recent review).
Here, $b\equiv (1-3w)/(1+3w)$ with $w$ being the Universe's equation of state at the emission time, $\Omega_{r,0}$ is the density fraction of 
radiation, $g_*(T)$ and $g_{*,s}(T)$ are the temperature-dependent
effective number of degrees of freedom for energy density and
entropy density, respectively, and $\mathcal{T}(u,v)$ is the transfer function 
\cite{Espinosa:2018eve,Kohri:2018awv}.
We denote with the subscript ``$H$'' 
the time when
induced GWs of the given wave number $k$
fall sufficiently within the Hubble horizon to behave as a radiation fluid in an expanding universe.

\begin{figure}[!t] 
    \centering
    \includegraphics[width=\columnwidth]{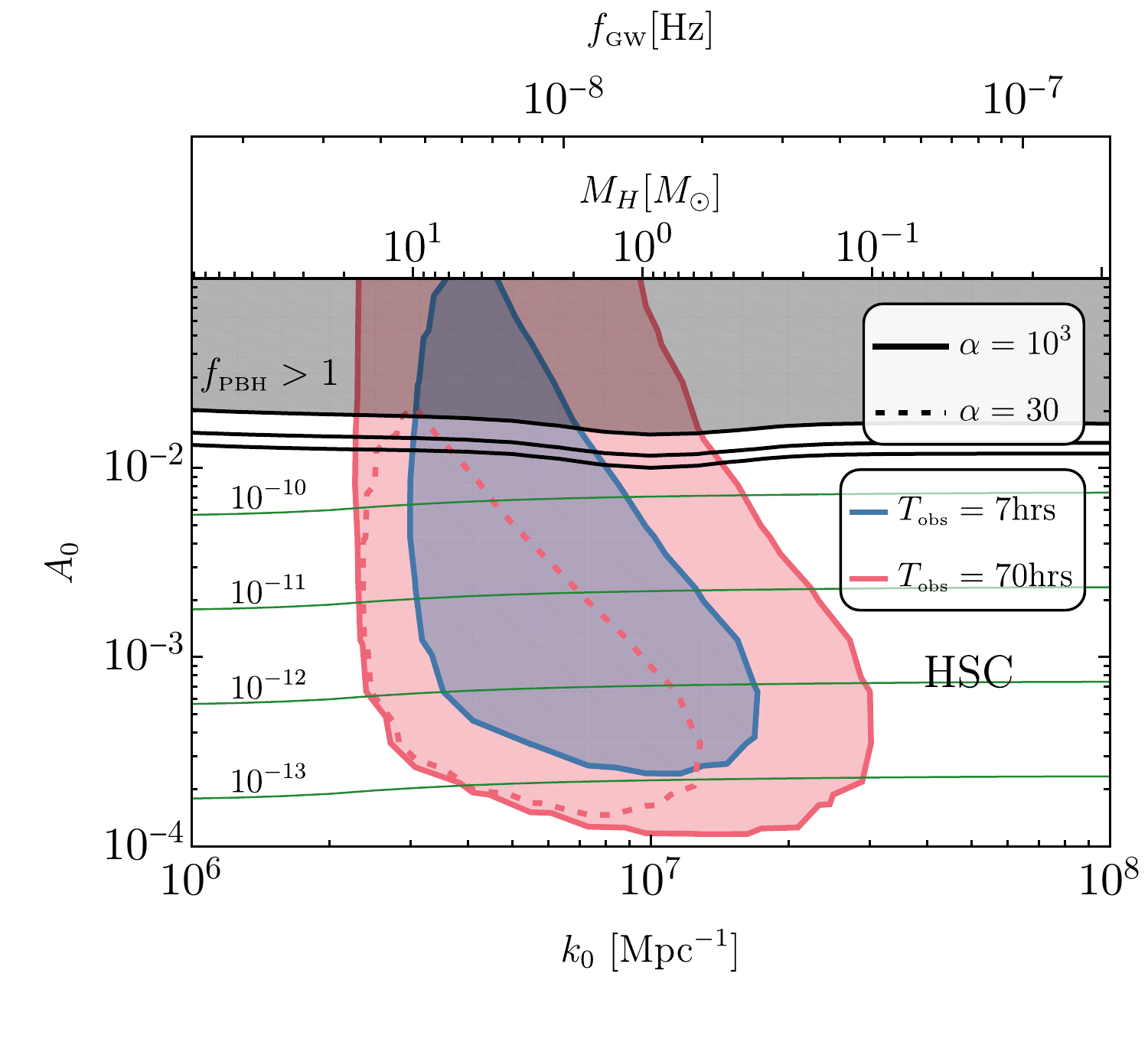}%
    \caption{
    Parameter space excluded by current (7~h) and future (70~h) HSC observations, 
    assuming a PL+Exp power spectrum of the form in Eq.~\eqref{eq:pzeta_PLexp} that peaks at amplitude $A_0$ at the scale wave number $k_0$.  
    The optimistic assumption $\alpha=10^3$ about the internal density of ultradense halos results in the blue region being ruled out currently and the larger red region in the forecast. For the more moderate assumption $\alpha = 30$, current data do not set constraints, but the 70-h observations are forecasted to rule out the region enclosed by the dashed line. If we assume $\alpha = 1$, then neither current nor the forecasted observations can set any constraint.
    At the top, we indicate the horizon mass $M_H$ corresponding to $k_0$ [assuming $g_* = 25$ and $\kappa = 2.5$; see Eq.~\eqref{M-k}], which is close to the mass scale of the PBHs that would result from primordial power at that scale.  
    We also indicate the SGWB peak frequency $f_\text{\tiny GW}$ corresponding to power at that scale. 
    We shade in gray the region already ruled out by requiring no overproduction of PBHs, while the solid black horizontal lines correspond to the PBH mass fractions $f_\text{\tiny PBH} = [1,10^{-3},10^{-5}]$ from top to bottom. 
    The horizontal green lines indicate the 
    energy density of the SGWB, $\Omega_\text{\tiny SGWB}=[10^{-10},10^{-11},10^{-12},10^{-13}]$.
}
    \label{fig:constraints_narrowPS}
\end{figure}

In Fig.~\ref{fig:constraints_narrowPS},
we show the GW abundance produced at second order by the curvature power spectrum in Eq.~\eqref{eq:pzeta_PLexp}.
Current and future constraints would be able to rule out the scalar-induced interpretation of the SGWB potentially hinted by the PTA data, again provided that the internal density of ultradense halos is sufficiently high ($\alpha\approx 10^3$) and that dark matter can cluster on the relevant $\sim 10^{7}$~Mpc$^{-1}$ scales.

\section{Conclusions and outlook}\label{sec:conclusions}

Constraining the primordial Universe is one of the fundamental endeavors of modern cosmology. 
While CMB and large-scale structure observations allow us to measure the amplitude of perturbations at megaparsec scales and higher, primordial fluctuations on smaller scales evade our observational capabilities.  
Developments in the theory of PBH formation and GW emission allow us to set conservative upper limits on the amplitudes of initial perturbations, while current and future GW data may allow for stronger constraints or, potentially, discoveries. 

PBHs and the stochastic GW background probe
scenarios where the amplitudes of perturbations are enhanced at small scales.
In this paper we have discussed a different probe of these scenarios: the potential lensing signatures left by the formation of ultradense dark matter halos in the early Universe, as modeled in Ref.~\cite{Delos:2022yhn}. 
Using data from the HSC, OGLE, and EROS surveys, we showed that it is possible to constrain the amplitude of perturbations at small scales. In particular, the Subaru-HSC observations may allow us to constrain primordial perturbations in a narrow range of scales around $k\sim 10^{7}$~Mpc$^{-1}$. Such scales correspond to the formation of stellar mass PBHs, which are of interest for LIGO/Virgo/Kagra and next-generation gravitational-wave detectors. Perturbations at these scales could also source a stochastic GW background in the nanohertz range, which is of potential interest to PTA experiments. Microlensing of ultradense halos could probe primordial spectral amplitudes as low as $\mathcal{P}_\zeta\simeq 10^{-4}$.

The principal limitation to this approach is that it requires that the dark matter be capable of clustering, and remaining clustered, at the relevant $k\sim 10^{7}$~Mpc$^{-1}$ scales. If the dark matter is too warm \cite{Bode:2000gq,Viel:2005qj} or too light \cite{Hu:2000ke,Li:2018kyk}, then its thermal motion or quantum pressure, respectively, could preclude the formation of the sub-Earth-mass halos that arise from perturbations on such scales.
However, particle dark matter heavier than about $10^{-7}$~eV \cite{Li:2018kyk} (including the QCD axion \cite{Marsh:2015xka}) can carry density perturbations at the relevant $\mathcal{O}(0.1)$~pc scales as long as it was never in kinetic equilibrium with the Standard Model plasma. So can particles that were in thermal contact with the plasma as long as they are heavier than $\mathcal{O}(100)$~GeV (with the precise threshold being sensitive to when kinetic decoupling occurred \cite{Green:2003un}).
\chg{Meanwhile, if the dark matter has significant nongravitational self-interactions, then collisional relaxation effects would likely suppress the density inside ultradense halos to a significant extent over cosmic time (e.g.~\cite{Tulin:2017ara}). A similar effect would arise from gravitational collisions if the dark matter is predominantly PBHs at mass scales not sufficiently smaller than the ultradense halo mass scale (e.g.~\cite{1987gady.book.....B}).
However, collisionless particle dark matter would preserve the internal density of ultradense halos.}
PBH dark matter toward the lower end of the asteroid-mass window (e.g.~\cite{Carr:2020gox}) is likely \chg{also light enough to maintain the density within} the relevant $10^{-8}$ to $10^{-7}$~M$_\odot$ halos.

Finally, we emphasize that our current and prospective constraints are subject to significant theoretical uncertainty regarding the properties of halos that form during the radiation epoch, particularly their internal density values. While optimistic assumptions produce very interesting constraints on primordial power that are relevant to the interpretations of ongoing GW and PTA experiments, conservative assumptions yield no constraint from microlensing. This outcome motivates more detailed explorations, likely with numerical simulations of cosmological volumes, of halo formation during the radiation epoch.


\begin{acknowledgments}
G.F. acknowledges financial support provided under the European
Union's H2020 ERC, Starting Grant Agreement No.~DarkGRA--757480 and under the MIUR PRIN program, and support from the Amaldi Research Center funded by the MIUR program ``Dipartimento di Eccellenza" (CUP:~B81I18001170001).
This work was supported by the EU Horizon 2020 Research and Innovation Program under the Marie Sklodowska-Curie Grant Agreement No. 101007855.
\end{acknowledgments}

\appendix

\section{Microlensing surveys}\label{app:Microsurv}

In this appendix, we summarize the setup of the three surveys we consider in this work. 
These are EROS-2~\cite{EROS-2:2006ryy}, 
OGLE-IV~\cite{Niikura:2019kqi} and Subaru-HSC~\cite{Niikura:2017zjd}.

\subsection{EROS}

The EROS-2 survey 
observes stars located within the Large Magellanic Cloud (LMC), which is placed at a distance $D_\text{\tiny S} = 50~{\rm kpc}$ away from Earth.
We neglect the contribution from 
Small Magellanic Cloud data
in this analysis as
its constraining power was shown to be subdominant compared to LMC sources. 
The lenses are assumed to be distributed within the Milky Way (MW), and we describe its dark matter density distribution as an isothermal profile~\cite{Cirelli:2010xx,Croon:2020wpr}
\begin{align}
\rho_{\rm DM} (r) &= \dfrac{\rho_s}{1+(r/r_{\rm iso})^2}
\end{align}
with $\rho_s=1.39\,{\rm GeV/cm^3}$ and $r_{\rm iso}=4.38\,{\rm kpc}$. The radial position $r$ can be rewritten from Earth's location as
\begin{equation}
r \equiv \sqrt{R^2_{\rm Sol} - 2 x R_{\rm Sol} D_\text{\tiny S} \cos \ell \cos b + x^2 D_\text{\tiny S}^2},
\label{eq:rdefsun}
\end{equation}
where
$R_{\rm sol}=8.5\,{\rm kpc}$ is the Sun's radial position
and 
$(\ell,b)=({280}^\circ,-{33}^\circ)$ are the LMC's sky coordinates.
The MW circular speed is taken to be approximately $v_0 = {\rm 220~km/s}$
\cite{2019ApJ...871..120E}.
The number of source stars that are used in the survey is  $N_\star = 5.49 \times 10^6$ and the observation time is $2500$ days. The  efficiency factor $\epsilon(\tE)$
can be found in  Fig.~11 of Ref.~\cite{EROS-2:2006ryy}.
As the EROS-2 LMC survey has only observed one candidate microlensing signature (which we assume to be of astrophysical origin), i.e. $N_{\rm obs}=1$, one can set an upper bound at $90\%$ confidence level by requiring $N_{\rm exp} \lesssim 3.9 $, assuming Poisson statistics. 

The constraint we derive adopting the EROS survey is shown in Fig.~\ref{fig:constraint_mono}, assuming a monochromatic mass distribution of lenses of various sizes, and occupies the range of masses $M\in[10^{-4} \divisionsymbol 10] M_\odot$.
As one can see, there is no difference in the constraint for lens density values $\bar\rho\gtrsim 10^{-9}$ M$_\odot$R$_\odot^{-3}$.
This is because such a lens is much smaller than its Einstein radius \eqref{eq:estRE} and close to the pointlike limit. A similar argument leads to the conclusion that the finite source size effect is also irrelevant. 
Therefore, in Eq.~\eqref{eq:Nevents} we marginalize the distribution of stellar sizes.

\subsection{OGLE}

The OGLE-IV survey adopts as light sources stars of the Milky Way bulge. 
When deriving the constraint based on the OGLE-IV survey, 
we describe the isothermal density profile of the MW halo as in the previous section.
However, we set the distance to the source stars as 
$D_\text{\tiny S} \simeq 8.5~{\rm kpc}$,  the longitude and latitude of the source in Galactic coordinates as $(\ell,b)=({1.09}^\circ,-{2.39}^\circ)$
and the detection efficiencies at the values provided in Ref.~\cite{Niikura:2019kqi}.
In this case, the number of source stars that are used in the survey is  $N_\star = 4.88 \times 10^7$
and the observation time is $1826$ days.

One important difference of OGLE-IV compared to the other surveys  is the presence of 2622 candidate events observed in their 5-year dataset, which is found to agree within 
 1\% with astrophysical models of standard foreground events~\cite{Niikura:2019kqi} (see also \cite{2023ApJ...944L..33C}). 
It is also interesting to notice that the survey identified $N_\text{\tiny obs} = 6$ events near $\tE \sim$ 0.1 days for which there is
no satisfactory explanation is found within the foreground model~\cite{2017Natur.548..183M,Niikura:2019kqi,Scholtz:2019csj} and that constitute potential PBH detections.
Here, we will assume it constitutes a foreground, regardless of its nature. 
We derive the constraint on the fraction $f$ of dark matter in lens objects by requiring that the combination \cite{Croon:2020wpr}
\begin{equation}\label{kappaeq}
\kappa = 2 \sum_{i = 1}^{N_{\rm bins}} \left[ N^{\rm FG}_i - N^{\rm SIG}_i + N^{\rm SIG}_i \ln \frac{N^{\rm SIG}_i}{N^{\rm FG}_i} \right]
\end{equation}
be smaller than $\kappa$ = 4.61, which corresponds to the 90\% confidence level assuming Poisson statistics. 
Here, the index $i$ indicates the 
binning of events by $\tE$ adopted in Ref.~\cite{Niikura:2019kqi}, $N^{\rm DM}_i$ 
is the number of lensing signals induced by dark matter halos from Eq.~\eqref{eq:Nevents},
$N^{\rm FG}_i$ is the number of astrophysical foreground events, and $N^{\rm SIG}_i \equiv N^{\rm FG}_i + N^{\rm DM}_i$.

The resulting constraint for a monochromatic halo mass distribution is shown in Fig.~\ref{fig:constraint_mono}. 
The OGLE survey is dominant in the range of masses 
$M\in[10^{-6} \divisionsymbol 10^{-3}] M_\odot$.
Due to the lighter lenses considered here, and the consequent smaller Einstein radius, the constraint begins to degrade when $\bar\rho\lesssim 10^{-9}$~M$_\odot$R$_\odot^{-3}$.

\subsection{HSC}

The Subaru-HSC survey~\cite{Niikura:2017zjd}
observed stars of the M31 galaxy, 
whose distances from us are approximately 
$D_\text{\tiny S}\simeq 770\,{\rm kpc}$.
The lensing signatures may arise 
from the presence of compact structures in both the MW and M31. 
The circular speeds are taken to be approximately $v_0 = {\rm 220~km/s}$ for the MW
\cite{2019ApJ...871..120E} and 
$v_0=250~{\rm km/s}$ for M31 \cite{Kafle:2018amm}.
The differential event rate is 
therefore the sum of two pieces
$   d\Gamma=d\Gamma_{\rm MW}+d\Gamma_{\rm M31}$.
We find that the contribution from lenses within the MW dominates the number of events. 
Following Ref~\cite{Niikura:2017zjd}, 
the spatial dark matter distribution of the MW 
is assumed to be given by an NFW profile
with scale density
$0.184\,{\rm GeV/cm^{3}}$,
scale radius
$21.5\,{\rm kpc}$, 
and $r$ determined from Earth's location as in Eq.~\eqref{eq:rdefsun} with  $(\ell,b)=(121.2^\circ,-21.6^\circ)$~\cite{Klypin:2001xu}.
The number of stars in the Subaru-HSC survey
is $N_\star=8.7\times 10^7$, while the observation time is $T_\text{\tiny obs}=7$~h.
Finally, the detection efficiency is given by Fig.~19 in Ref.~\cite{Niikura:2017zjd}.
Following Ref.~\cite{Croon:2020ouk}, we approximate the detection efficiency as $\epsilon =0.5$ in the regime with
$2\,{\rm min}\leq t_\text{\tiny E} \leq 7\,{\rm h}$.

The constrained lens masses in the Subaru-HSC survey are much lighter than in the case of EROS and OGLE surveys. 
Therefore, they correspond to much smaller 
Einstein radii for which both extended lens size and
the finite source size corrections are important.
In particular, the latter effect was neglected in the first version of the analysis of Ref.~\cite{Niikura:2017zjd} that lead to an overestimation of the number of detectable events below around $10^{-11} M_\odot$, which is instead drastically suppressed. 
Following Ref.~\cite{Smyth:2019whb}, when integrating over the source star radii of M31 $R_\star$ in Eq.~\eqref{eq:Nevents}, 
we adopt the distribution derived using the Panchromatic Hubble Andromeda Treasury star catalog~\cite{2014ApJS..215....9W,2012ApJS..200...18D} and the MESA Isochrones and Stellar Tracks stellar evolution package~\cite{2016ApJ...823..102C,2016ApJS..222....8D}
(see Fig.~4 of Ref.~\cite{Smyth:2019whb} and related discussion for more details).
Also in this case, we consider the HSC single microlensing event candidate as foreground and require $N_{\rm exp} \lesssim 3.9$, which corresponds to the 90\% confidence level assuming Poisson statistics.

\bibliography{main}

\end{document}